\documentclass[11pt]{amsart}
\usepackage[margin=1in]{geometry}
\usepackage{amsmath,amssymb,amsthm,mathtools}
\usepackage[expansion=false]{microtype}
\usepackage{enumitem}
\usepackage{tikz}
\usepackage{mathpazo}
\usetikzlibrary{arrows.meta,decorations.pathreplacing}
\usepackage[colorlinks=true,linkcolor=blue,citecolor=blue,urlcolor=blue]{hyperref}

\numberwithin{equation}{section}
\renewcommand{\S}{\mbox{$\mathsection$}}
\newtheorem{theorem}{Theorem}[section]
\newtheorem{proposition}[theorem]{Proposition}
\newtheorem{lemma}[theorem]{Lemma}
\newtheorem{corollary}[theorem]{Corollary}
\theoremstyle{definition}

\theoremstyle{remark}
\newtheorem{remark}[theorem]{Remark}

\newcommand{\C}{\mathbb{C}}
\newcommand{\A}{\mathbb{A}}
\newcommand{\g}{\mathfrak{g}}
\newcommand{\sltwo}{\mathfrak{sl}_2}
\newcommand{\hh}{\mathfrak{h}}
\newcommand{\tl}{\mathfrak{t}}
\newcommand{\lie}{\mathfrak{l}}
\newcommand{\pp}{\mathfrak{p}}
\newcommand{\dd}{\partial}
\newcommand{\Obs}{\mathsf{Obs}_{\mathsf{HT}}}
\newcommand{\cC}{\mathcal{C}}
\newcommand{\cO}{\mathcal{O}}
\newcommand{\cK}{\mathcal{K}}
\newcommand{\cA}{\mathcal{A}}
\newcommand{\Xc}{\mathsf{X}}
\newcommand{\Yc}{\mathsf{Y}}
\newcommand{\Sym}{\operatorname{Sym}}
\newcommand{\Spec}{\operatorname{Spec}}
\newcommand{\Supp}{\operatorname{Supp}}
\newcommand{\Res}{\operatorname{Res}}
\newcommand{\depth}{\operatorname{depth}}
\newcommand{\im}{\operatorname{im}}
\newcommand{\Hom}{\operatorname{Hom}}
\newcommand{\sgn}{\operatorname{sgn}}
\newcommand{\ev}{\operatorname{ev}}
\newcommand{\gh}{\mathrm{gh}}
\newcommand{\wt}{\mathrm{wt}}
\newcommand{\Hb}{\overline{H}}
\newcommand{\Car}{\mathrm{C}}
\newcommand{\bO}{\mathbf{O}}
\newcommand{\bK}{\mathbf{K}}
\newcommand{\Gr}{\mathrm{Gr}}
\newcommand{\Ext}{\operatorname{Ext}}

\title{Lie Algebra Cohomology and Seiberg-Witten Theory}
\author{Ahsan Z. Khan}
\address{Harvard University, Center of Mathematical Sciences and Applications, Cambridge, MA 02138}
\email{ahsankhan@g.harvard.edu}
\date{\today}

\begin{document}

\begin{abstract}
We compute the algebra of perturbative holomorphic-topological local operators of pure $\mathcal N=2$ gauge theory with gauge group $SU(2)$, proving a conjecture of the author: it is the Poisson vertex algebra generated by a Virasoro element $X$ of central charge zero and an odd primary $Y$ of weight three, subject only to $X^2=0$. Equivalently, we compute the relative Lie algebra cohomology $H^*(\mathfrak{sl}_2[z,w]/(w^2),\mathfrak{sl}_2)$ and the self-Ext groups of the unit in the coherent Satake category for $SL_2$ and $PGL_2$. The proof uses factorization on configuration spaces of points on the affine line, a depth argument, and two deformations of the coefficient algebra $\mathbb C[z,w]/(w^2)$.
\end{abstract}

\maketitle

\section{Introduction}

Much of the appeal of supersymmetric quantum field theories comes from the ability to calculate non-trivial quantities exactly. Examples include the Witten index, the elliptic genus, the superconformal index and the Schur index. A deeper study often reveals that these indices are just the shadow of a much richer invariant, carrying interesting and non-trivial algebraic structure. For example, the Schur index of a four-dimensional $\mathcal N=2$ superconformal field theory is the shadow of a chiral algebra \cite{BLLPRR15}. Studying the spaces underlying these indices also reveals that such ``superconformal'' indices are in fact well-defined in the more general non-conformal setting: they are characters of the space of local operators in a mixed holomorphic-topological twist \cite{Kap06,Kha26}.

This paper studies the space of local operators in the holomorphic-topological twist of the pure four-dimensional $\mathcal{N}=2$ gauge theory with gauge group $SU(2)$ \cite{SW94}. In a previous paper \cite{Kha26}, the author proposed the following conjecture: to all orders in perturbation theory, the space of local operators is given by the Poisson vertex algebra $A$, generated by an even field $X$ and an odd field $Y$ with $\lambda$-brackets
\begin{equation}\label{eq:brackets}
\{X_\lambda X\}=\dd X+2\lambda X,\qquad
\{X_\lambda Y\}=\dd Y+3\lambda Y,\qquad
\{Y_\lambda Y\}=0,
\end{equation}
divided by the smallest Poisson vertex ideal containing $X^2$. In this paper we prove that conjecture.

To state the result more precisely, we recall from \cite{Kha26} that the space of local operators of the twisted $SU(2)$ theory is given, to all orders in perturbation theory \cite{ESW22,BG25}, by the relative Lie algebra cohomology
\begin{equation}\label{eq:intro-obs}
\Obs(\sltwo)
=H^*\bigl(\sltwo[z,w]/(w^2),\sltwo\bigr),
\end{equation}
where $\sltwo[z,w]/(w^2)$ denotes the infinite-dimensional Lie algebra $\sltwo\otimes\C[z,w]/(w^2)$ and $w$ denotes an even square-zero variable (written $\varepsilon=dz$ in \cite{Kha26}). There we showed that this cohomology carries a Poisson vertex algebra structure, and constructed a homomorphism of Poisson vertex algebras $\varphi: A \to \Obs(\sltwo)$.

\begin{theorem}\label{thm:main}
The map $\varphi:A\to\Obs(\sltwo)$ is an isomorphism of Poisson vertex algebras. In particular,
\begin{equation}\label{eq:main-presentation}
H^*\bigl(\sltwo[z,w]/(w^2),\sltwo\bigr)\;\cong\;
\frac{\C[X,\dd X,\ldots]\otimes\Lambda[Y,\dd Y,\ldots]}{\langle X^2,\,XY,\,X\dd Y,\,Y\dd Y\rangle_\dd}
\end{equation}
as differential supercommutative algebras graded by spin, ghost number and $B$-number.
\end{theorem}

Thus every perturbative local operator of the twisted $\mathcal N=2$ theory is a differential polynomial in the two gaugino composites
\begin{equation}\label{eq:gaugino}
X=-\operatorname{tr}\bigl(\bar\lambda\lambda\bigr),\qquad
Y=\tfrac13\operatorname{tr}\bigl(\bar\lambda^3\bigr),
\end{equation}
where $\lambda$ and $\bar\lambda$ are the gaugino components, one of each chirality, that survive the holomorphic-topological twist,\footnote{In terms of the ghosts of \S\ref{subsec:basic}, $X=-\operatorname{tr}(b\,\dd c)$ and $Y=\frac13\operatorname{tr}(b^3)$.} and all relations among them follow from $X^2=0$ and the axioms of a Poisson vertex algebra.

Recall that $\Obs(\sltwo)$ is graded by ghost number, spin and ``$B$-number'', which in terms of~\eqref{eq:intro-obs} are the cohomological degree and the gradings induced by $\deg z=\deg w=-1$ and by $\deg z=0$, $\deg w=-1$, respectively. Each trigraded piece is finite-dimensional. Theorem~\ref{thm:main} determines these dimensions, and combined with the computation of the Hilbert--Poincar\'e series of $A$ in \cite[(2.65)]{Kha26}, it gives the following.

\begin{corollary}\label{cor:intro}
Let $t$, $q$ and $y$ keep track of the ghost number, the spin and the $B$-number. Then
\begin{equation}\label{eq:intro-poincare}
P_{\Obs(\sltwo)}(t,q,y)=\sum_{n\ge0}q^{n(n+1)}\,t^{2n}y^{n}\,\frac{(-tqy^2;q)_n}{(q;q)_n}.
\end{equation}
\end{corollary}
Specializing to $t=-1$ gives the Macdonald index of the pure $\mathcal{N}=2$ theory,
\begin{equation}\label{eq:intro-Mac}
\mathcal I_{\mathrm M}(q,y)=\sum_{n\ge0}q^{n(n+1)}y^n\,\frac{(qy^2;q)_n}{(q;q)_n},
\end{equation}
and further setting $y=1$ gives its Schur index,
\begin{equation}\label{eq:intro-Schur}
\mathcal I_{\mathrm S}(q)=\sum_{n\ge0}q^{n(n+1)}.
\end{equation}
To our knowledge, the formula~\eqref{eq:intro-Mac} for the Macdonald index of the pure $SU(2)$ theory is new.\footnote{In \cite{Son16}, a different expression for the Macdonald index was proposed on the basis of a 2d TQFT description of the theory, and checked to low order in $q$.}

The Macdonald index is also given by a Molien-type integral over the maximal torus of $SU(2)$. Comparing this integral with~\eqref{eq:intro-Mac} proves a new constant term identity, predicted in \cite[(3.127)]{Kha26}:\footnote{Here $\operatorname{CT}_z$ denotes the constant term in $z$, and $(qx^{\pm1};q)_\infty:=(qx;q)_\infty(qx^{-1};q)_\infty$.}
\begin{equation}\label{eq:intro-CT}
\operatorname{CT}_z\Bigl[\tfrac12(1-z^2)(1-z^{-2})\,(q;q)_\infty(qz^{\pm2};q)_\infty\,(qy;q)_\infty(qyz^{\pm2};q)_\infty\Bigr]
=\sum_{n\ge0}q^{n(n+1)}y^n\,\frac{(qy^2;q)_n}{(q;q)_n}.
\end{equation}

\subsection{Outline of the Proof}\label{subsec:idea}

Injectivity of $\varphi$ is the easier half of Theorem~\ref{thm:main}. In \S\ref{sec:injective} we construct, for every element of the graded dual of $A$, an explicit Chevalley--Eilenberg cycle that detects it. The cycles live on the Cartan current algebra $\hh[z]\ltimes w\,\sltwo[z]$, where the chain complex becomes a Koszul complex, and a change of variables from power sums to elementary symmetric functions produces them in closed form.

The main difficulty is surjectivity: one has to show that the relative cohomology contains nothing beyond $\varphi(A)$. Our argument is algebro-geometric and goes as follows.

Fix the $B$-number $B$, that is, the number of $b$-ghosts. The complex $\text{Obs}_{\text{HT}}(\mathfrak{sl}_2)$ comes equipped with derivations $T_m$, $m\ge1$, which lower the mode of every $b$-ghost by $m$, annihilate the $c$-ghosts and commute with the BRST differential. On the graded dual of the cohomology they act as the power sums $\sum_it_i^m$ of $B$ commuting variables $t_1,\dots,t_B$, which we think of as the ``positions'' of the $b$-ghosts on the $z$-line. In this way the dual of the cohomology at $B$-number $B$ becomes a coherent sheaf $\cK_B$, graded by the number $p$ of $c$-ghosts, on the configuration space
\begin{equation}\label{eq:intro-symB}
\Sym^B\A^1=\{\text{effective divisors } t_1+\dots+t_B \text{ on the } z\text{-line}\}\;\cong\;\A^B .
\end{equation}
The relative BGG resolution presents $\cK_B$ as the homology of a complex of vector bundles of length $B$ on $\Sym^B\A^1$ (\S\ref{sec:sheaves}).

\begin{figure}[t]
\centering
\begin{tikzpicture}[>=Stealth, every node/.style={font=\small}]
\begin{scope}
  \draw[->] (0,0) -- (4.2,0) node[right] {$z$};
  \foreach \x in {0.5,1.4,3.5} {\fill (\x,0) circle (2.2pt); \node[below=4pt] at (\x,0) {$X$};}
  \fill (2.45,0) circle (2.2pt);
  \draw (2.45,0) circle (4.6pt);
  \draw (2.45,0) circle (7pt);
  \node[below=8pt] at (2.45,0) {$Y$};
  \node at (2.1,-1.75) {(a) ghosts as points};
\end{scope}
\begin{scope}[xshift=5.4cm]
  \draw[->] (0,0.55) -- (4.2,0.55) node[right] {$z$};
  \foreach \x in {0.4,0.8,1.2} \fill (\x,0.55) circle (2pt);
  \foreach \x in {2.85,3.25,3.65} \fill (\x,0.55) circle (2pt);
  \draw[decorate,decoration={brace,amplitude=4pt}] (0.28,0.75) -- (1.32,0.75) node[midway,above=4pt] {$D_1$};
  \draw[decorate,decoration={brace,amplitude=4pt}] (2.73,0.75) -- (3.77,0.75) node[midway,above=4pt] {$D_2$};
  \draw[->] (0,-0.6) -- (4.2,-0.6) node[right] {$z$};
  \fill (2.05,-0.6) circle (2pt);
  \draw (2.05,-0.6) circle (4pt);
  \draw (2.05,-0.6) circle (6pt);
  \draw (2.05,-0.6) circle (8pt);
  \node[below=9pt] at (2.05,-0.6) {\scriptsize $B\cdot a\in Z_B$};
  \node at (2.1,-1.75) {(b) separation and collision};
\end{scope}
\begin{scope}[xshift=10.9cm]
  \draw[->] (-0.3,0) -- (3.5,0) node[right] {$z$};
  \draw[->] (0,-1.2) -- (0,1.35) node[above] {$w$};
  \draw[blue!70!black, thick, domain=-1.1:1.1, smooth, variable=\t] plot ({2.3*\t*\t}, {\t});
  \draw[red!70!black, thick, dashed] (-0.2,0.72) -- (3.2,0.72);
  \draw[red!70!black, thick, dashed] (-0.2,-0.72) -- (3.2,-0.72);
  \draw[line width=2.4pt] (0,0) -- (3.2,0);
  \node[above] at (1.45,0.02) {\scriptsize $w^2=0$};
  \node[blue!70!black, right] at (2.8,1.1) {\scriptsize $w^2=z$};
  \node[red!70!black, right] at (3.2,-0.72) {\scriptsize $w^2=1$};
  \node at (1.6,-1.75) {(c) deformation};
\end{scope}
\end{tikzpicture}
\caption{The three steps of the proof. (a) A cohomology class of $B$-number $B$ is a function of the positions of $B$ ghosts on the $z$-line. The generator $X$ is a simple point and $Y$ is a triple point. (b) When the ghosts form separated clusters the cohomology factorizes, so any defect is supported on the small diagonal $Z_B$, where all ghosts collide. There it is removed by a depth argument. (c) In the two lowest degrees the double line $w^2=0$ is deformed into the parabola $w^2=z$ and into the pair of lines $w^2=1$.}
\label{fig:idea}
\end{figure}
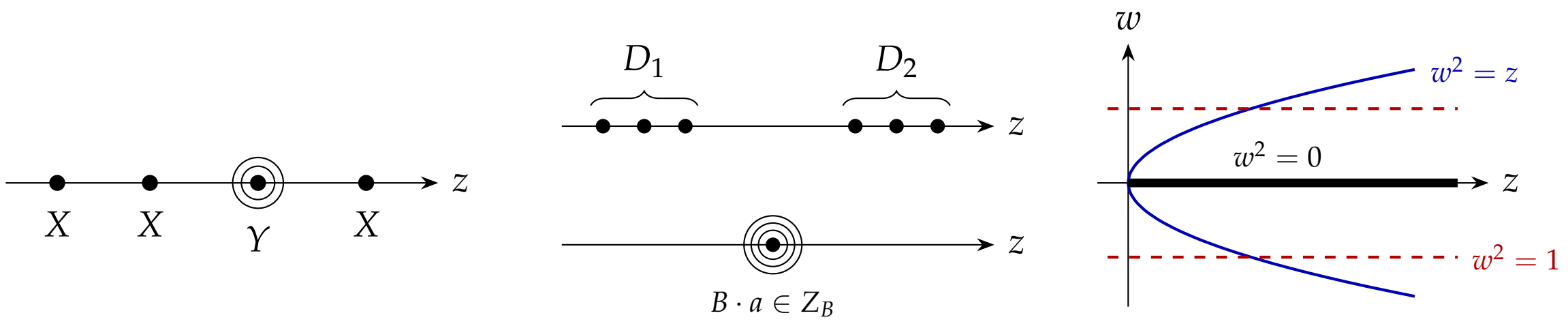

The algebra $A$ defines a second sheaf on the same space. The generator $X\mapsto-\delta_{ab}\,b^a\dd c^b$ carries a single $b$-ghost: it is a ``simple point''. The generator $Y\mapsto b^1b^2b^3$ carries three $b$-ghosts at the same location: it is a ``triple point''. Accordingly, the part of $A$ with $r$ factors of $X$ and $s$ factors of $Y$ is dual to the pushforward of the structure sheaf along the finite map
\begin{equation}\label{eq:intro-nu}
\nu_{r,s}:\Sym^r\A^1\times\Sym^s\A^1\longrightarrow\Sym^{r+3s}\A^1,\qquad (D,D')\longmapsto D+3D',
\end{equation}
which is the normalization of the closure of the stratum of divisors with $r$ simple and $s$ triple points. The relations $X^2=XY=X\dd Y=Y\dd Y=0$ say that dual elements, as functions of the positions, vanish to second order whenever two points collide, and to third order when both are triple points (Lemma~\ref{lem:dualA}). Injectivity of $\varphi$ gives a surjection of sheaves
\begin{equation}\label{eq:intro-psi}
\psi_B:\cK_B\twoheadrightarrow\cA^\vee_B:=\bigoplus_{r+3s=B}\nu_{r,s*}\cO,
\end{equation}
and Theorem~\ref{thm:main} asserts that its kernel, the ``defect'', vanishes. We prove this in three steps, which we call ``factorization'', ``depth'' and ``deformation''. See Figure~\ref{fig:idea}.

\noindent\emph{Factorization.}
Over the open subset of $\Sym^B\A^1$ where a divisor splits as $D_1+D_2$ with disjoint supports, both sheaves factorize:
\begin{equation}\label{eq:intro-factor}
\cK_B\cong\cK_{B_1}\boxtimes\cK_{B_2},\qquad
\cA^\vee_B\cong\cA^\vee_{B_1}\boxtimes\cA^\vee_{B_2}.
\end{equation}
For $\cA^\vee$ this is the Chinese remainder theorem. For $\cK$ it is a Lie-algebraic form of locality: the cohomology of $\sltwo[z]$ with coefficients supported at two disjoint finite sets of points is the tensor product of the two cohomologies. We deduce this from Teleman's theorem by an explicit cocycle computation (\S\ref{sec:separation}). In fact, \eqref{eq:intro-factor} is part of a richer structure: both sheaves are factorization coalgebras on the line, in the spirit of \cite{BD04}, and $\psi$ is a morphism between them (Remark~\ref{rem:factorization}). From this point of view, Theorem~\ref{thm:main} simply says that $\cK$ is the factorization coalgebra cogenerated by a simple point and a triple point, with the collision orders above (Remark~\ref{rem:dualA}).\footnote{ For the proof, however, we only need~\eqref{eq:intro-factor} as abstract isomorphisms.} By induction on $B$, the map $\psi_B$ is then an isomorphism away from the \emph{small diagonal}
\begin{equation}\label{eq:intro-ZB}
Z_B=\{B\cdot a:\ a\in\A^1\}\subset\Sym^B\A^1,
\end{equation}
where all $B$ ghosts collide. Thus the defect is supported on $Z_B$, a line of codimension $B-1$.

\noindent\emph{Depth.}
By the acyclicity lemma of Peskine and Szpiro \cite{PS73} (see also \cite{BE73}), the homology of a short complex of vector bundles cannot contain a nonzero subsheaf supported in high codimension, provided the quotient by it is sufficiently deep there. The top term of the BGG complex, which corresponds to the pure-$X$ sector, is a line bundle and splits off by a rank argument. What remains has length $B-1$, exactly the codimension of $Z_B$, while $\cA^\vee_B$ is deep along $Z_B$: the preimage of $Z_B$ under $\nu_{r,s}$ is a line in the $(r+s)$-dimensional space $\Sym^r\A^1\times\Sym^s\A^1$, so its codimension $r+s-1$ is at least the number $p=r$ of $c$-ghosts as soon as $s\ge1$. A depth count then forces the defect to vanish in every degree $p\ge2$ (\S\ref{sec:collision}).

\noindent\emph{Deformation.}
The depth bounds says nothing about the two lowest degrees $p=0,1$. There we deform the coefficient algebra: in the $(z,w)$-plane, the double line $w^2=0$ deforms into the parabola $w^2=z$ and into the pair of lines $w^2=1$,
\begin{equation}\label{eq:intro-deform}
\C[z,w]/(w^2-z)\cong\C[u],\qquad\C[z,w]/(w^2-1)\cong\C[z]\times\C[z].
\end{equation}
For both deformed algebras the relative cohomology follows from the Garland--Lepowsky theorem and it is tiny:  $\C$ and $\C\oplus\C\,Y$, respectively. The corresponding spectral sequences show that an excess class in degree $p=1$ and $B$-number $B$ would have to cancel against an excess class in degree $p=0$ and $B$-number $B+2$, through a map which raises the spin by one for the parabola and by two for the pair of lines. At the lowest spin carrying an excess class these two requirements are incompatible (\S\ref{sec:deformation}). The algebra $A$ does not interfere, since both deformations act on it with explicit contracting homotopies.

Combining the three steps proves Theorem~\ref{thm:main} by induction on $B$ (\S\ref{sec:proof}).

\subsection{Relation to the Affine Grassmannian}\label{subsec:Gr}
There is a second, more geometric way to think about Theorem~\ref{thm:main}, in terms of coherent sheaves on the affine Grassmannian. Let $G$ be $SL_2$ or $PGL_2$, let $\bO=\C[[z]]$ and $\bK=\C((z))$, and let $\Gr_G=G(\bK)/G(\bO)$ be the affine Grassmannian. Following a proposal of Costello reported in \cite{CW19}, the category of line operators in the holomorphic-topological twist \cite{Kap06} of pure $\mathcal N=2$ gauge theory is the derived category of $G(\bO)$-equivariant coherent sheaves on $\Gr_G$, made monoidal by convolution, which models the collision of line operators \cite{KS09}. With respect to the perverse coherent t-structure \cite{AB10,BFM05}, its heart is the \emph{coherent Satake category} of Cautis and Williams \cite{CW19}. Unlike its constructible counterpart in the geometric Satake correspondence \cite{MV07}, this category is neither semisimple nor symmetric monoidal. Its simple objects correspond to Wilson--'t~Hooft lines, its Grothendieck ring is the quantized K-theoretic Coulomb branch of the theory \cite{BFM05,BFN18}, and for $GL_n$ it categorifies a quantum cluster algebra \cite{CW19}. The construction extends to gauge theories with matter \cite{CW23}.

In any such description, bulk local operators are the derived endomorphisms of the identity line operator, that is, of the monoidal unit, which is the skyscraper sheaf $\cO_e$ at the base point $e=[G(\bO)]$. Niu identified this endomorphism complex with the basic $bc$-complex \cite[Theorem~4.13]{Niu22}. The tangent space of $\Gr_G$ at $e$ is
\begin{equation}\label{eq:tangent}
T_e\Gr_G=\g(\bK)/\g(\bO)\cong\bigl(\g(\bO)\,dz\bigr)^\vee,
\end{equation}
where the isomorphism is the residue pairing. Thus the self-Ext of $\cO_e$ on the formal neighborhood of $e$ is the exterior algebra $\Lambda^\bullet(\g(\bK)/\g(\bO))$, and passing to derived $G(\bO)$-invariants adds the Chevalley--Eilenberg cochains of $z\g(\bO)$. The result is the basic $bc$-complex: the $b$-ghosts become tangent vectors to $\Gr_G$ at $e$, and the $c$-ghosts become the generators of the $G(\bO)$-equivariant cohomology. Niu showed that the graded Euler characteristic of this complex is the Schur index \cite{Niu22}, but did not determine its cohomology. Since the complex depends on $G$ only through $\g=\sltwo$, Theorem~\ref{thm:main} computes this cohomology, and with it the derived endomorphisms of the unit of the coherent Satake category, for both $SL_2$ and $PGL_2$. Here we use the model of \cite[\S4]{Niu22}: sheaves are ind-coherent and equivariant for $\widetilde G(\bO)=G(\bO)\rtimes\C^\times$, where $\C^\times$ acts through the double cover of loop rotation, and $\Ext^\bullet_{\widetilde G(\bO)}$ denotes the graded $\Ext$, that is, the direct sum of the $\Ext$ groups into all twists of the target by characters of $\C^\times$, graded by the corresponding weight of loop rotation.

\begin{corollary}\label{cor:Gr}
Let $G=SL_2$ or $PGL_2$. There is an isomorphism of trigraded vector spaces
\begin{equation}\label{eq:Gr-ext}
\Ext^\bullet_{\widetilde G(\bO)}(\cO_e,\cO_e)\;\cong\;H^*\bigl(\sltwo[z,w]/(w^2),\sltwo\bigr)\;\cong\;A,
\end{equation}
under which the $\Ext$ degree, the polyvector degree and the weight of loop rotation correspond to the ghost number, the $B$-number and the spin, respectively. In particular, the Poincar\'e series of $\Ext^\bullet_{\widetilde G(\bO)}(\cO_e,\cO_e)$ is given by Corollary~\ref{cor:intro}.
\end{corollary}

Under this identification, $Y$ is the invariant trivector on the first-order normal directions $\g\,z^{-1}\subset T_e\Gr_G$, and $X$ pairs these directions with the first-order directions $z\g$ of $G(\bO)$. Moreover, the derivations $T_m$ of \S\ref{subsec:idea} act on $T_e\Gr_G$ as multiplication by $z^m$, so the factorization of \S\ref{sec:separation} is a Lie-algebraic shadow of the factorization of the Beilinson--Drinfeld Grassmannian over configurations of points \cite{BD04,MV07}.

Corollary \ref{cor:Gr} is an isomorphism of graded vector spaces. We expect that it can be upgraded to an isomorphism of $(E_2 \,, \, \text{chiral})$ algebras. 

Seen as a statement about the coherent Satake category, Theorem~\ref{thm:main} is perhaps surprising. In the constructible equivariant Satake category, the derived endomorphism algebra of the unit is just $H^*_{G}(\mathrm{pt})$, a polynomial ring (cf.\ \cite{BF08}). In the coherent setting it is instead the algebra $A$: infinitely generated, by the $\dd^kX$ and $\dd^kY$, with all relations following from $X^2=0$. Since the endomorphisms of the unit of a monoidal category act on all of its morphism spaces, $A$ also acts on the spaces of junctions between line operators, such as those between Wilson and 't~Hooft lines. Wilson--'t~Hooft lines and their junctions will be the subject of a subsequent paper.

\subsection{Higher Rank}\label{subsec:higher-intro}
It is natural to ask what happens for gauge group $SU(N)$, where the local operators form the analogue $\Obs(\mathfrak{sl}_N)$ of~\eqref{eq:intro-obs}. Writing $b$ and $c$ for the $\mathfrak{sl}_N$-valued ghost fields, the natural candidates for generators are the single-trace operators
\begin{equation}\label{eq:intro-single-trace}
X^{(i)}=-\operatorname{tr}\bigl(b^{2i+1}\dd c\bigr)\quad(i\ge0),\qquad Y^{(i)}=\tfrac1{2i+1}\operatorname{tr}\bigl(b^{2i+1}\bigr)\quad(i\ge1),
\end{equation}
which include $X=X^{(0)}$ and $Y=Y^{(1)}$. In the limit $N\to\infty$ these are in fact all there is: by reducing to single-trace operators, as in the theorem of Loday--Quillen and Tsygan, we show that the local operators are freely generated by~\eqref{eq:intro-single-trace} as a differential graded-commutative algebra, with $\lambda$-brackets~\eqref{eq:XX-stable}--\eqref{eq:YY-stable}. For finite $N\ge3$ we are not aware of a compact complete set of relations. We do, however, prove that the classical stress tensor $X$ is nilpotent, generalizing the relation $X^2=0$:

\begin{theorem}\label{thm:intro-XN}
Let $N\ge2$. In $\Obs(\mathfrak{sl}_N)$,
\begin{equation}\label{eq:intro-XN}
X^N=0\qquad\text{and}\qquad X^{N-1}\ne0.
\end{equation}
\end{theorem}

Interestingly, this parallels the classical relation $S^N=0$ for the glueball superfield in the chiral ring of pure $\mathcal N=1$ super Yang--Mills theory \cite{CDSW02}.

\subsection{Relation to Other Work}
The cohomology of current algebras with nilpotent coefficients has a long history. The strong Macdonald conjecture of Hanlon and Feigin \cite{Han90,Fei88}, proved by Fishel, Grojnowski and Teleman \cite{FGT08}, determines the cohomology of the truncated current algebras $\g[z]/(z^N)$ and of the current superalgebra $\g[z,s]$ in which $s$ is an \emph{odd} square-zero variable. The Lie algebra $\sltwo[z,w]/(w^2)$ differs from $\sltwo[z,s]$ only in the parity of the square-zero variable, so it is natural to ask whether these techniques apply to it. The answer, however, is of a different nature: the cohomology rings computed in \cite{FGT08} are free, whereas the algebra $A$ has infinitely many relations. Accordingly, our method is different: instead of harmonic representatives, we use the module structure over symmetric functions, factorization and depth. Factorization at disjoint points of the cohomology of current algebras is also a central theme of Teleman's work on Lie algebra cohomology and the fusion rules \cite{Tel95,Tel96}.

\subsection{Organization of the Paper}
The paper is organized as follows. In \S\ref{sec:setup} we recall the algebra $A$, the basic complex and the map $\varphi$, and reformulate Theorem~\ref{thm:main} in dual form. Section~\ref{sec:inputs} collects the representation-theoretic input. Section~\ref{sec:injective} proves injectivity. Section~\ref{sec:sheaves} sets up the sheaves on $\Sym^B\A^1$ and treats the top degree. Sections~\ref{sec:separation}, \ref{sec:collision} and~\ref{sec:deformation} carry out separation, collision and deformation, respectively, and \S\ref{sec:proof} assembles the proof and derives the corollaries. Finally, in \S\ref{sec:higher} we determine the limit of $H^*(\mathfrak{sl}_N[z,w]/(w^2),\mathfrak{sl}_N)$ as $N\to\infty$, as a Poisson vertex algebra, and prove the relation $X^N=0$ for finite $N$.

\subsection{Notation and Conventions}\label{subsec:conventions}
All vector spaces, algebras and Lie algebras are over $\C$. For a commutative $\C$-algebra $R$ and a Lie algebra $\g$ we write $\g\otimes R$ for the current algebra, and for an ideal $(f)\subset\C[z,w]$ we abbreviate $\g[z,w]/(f):=\g\otimes\C[z,w]/(f)$. For a graded vector space $W=\bigoplus_SW_S$ with finite-dimensional pieces we write $W^\vee=\bigoplus_SW_S^*$ for its graded dual. The conventions for cochains of current algebras, polynomial in the modes or with finite-dimensional coefficients, are fixed in \S\ref{subsec:basic} and at the beginning of \S\ref{sec:inputs}.

\subsection{Formalization and Use of AI Tools}
Several of the proof techniques in this paper were suggested by the large language models GPT-6 Astra and Claude Opus 5.5. Claude Opus 5.5 also helped with the writing and proofreading of the manuscript. Under the author's direction, the two models were also coordinated to produce a complete Lean formalization of Theorem~\ref{thm:main}, available in a companion repository.\footnote{\url{https://github.com/qqz3ro/pure-su2-pva}} The author takes full responsibility for the content of this paper, including any errors.

\section*{Acknowledgements}
I thank Bowen Yang for stimulating discussions on commutative algebra, and Kevin Costello for many illuminating discussions on holomorphic-topological twists and related topics. This work is supported by the Center of Mathematical Sciences and Applications at Harvard University.

\section{The Algebra \texorpdfstring{$A$}{A} and the Basic Complex}\label{sec:setup}

We first describe the two sides of Theorem~\ref{thm:main}: the Poisson vertex algebra $A$, and the relative Lie algebra cohomology $\Obs(\sltwo)$ together with the complex that computes it. We then recall the map $\varphi$ between them.

\subsection{The Poisson Vertex Algebra \texorpdfstring{$A$}{A}}\label{subsec:A}
Recall that a \emph{Poisson vertex algebra} \cite{BDK09} is a supercommutative algebra $V$ with an even derivation $\dd$ and a $\lambda$-bracket $V\otimes V\to V[\lambda]$, $a\otimes b\mapsto\{a_\lambda b\}$, which makes $(V,\dd)$ a Lie conformal superalgebra \cite{Kac98} and acts by derivations:
\begin{equation}\label{eq:leibniz}
\{a_\lambda bc\}=\{a_\lambda b\}c+(-1)^{|a||b|}\,b\{a_\lambda c\}.
\end{equation}
A \emph{Poisson vertex ideal} is a differential ideal $I\subset V$ with $\{V_\lambda I\}\subset I[\lambda]$. The quotient $V/I$ is again a Poisson vertex algebra.

Let $R=\C[\dd]X\oplus\C[\dd]Y$ be the Lie conformal superalgebra with $X$ even, $Y$ odd and $\lambda$-brackets
\begin{equation}\label{eq:R-brackets}
\{X_\lambda X\}=\dd X+2\lambda X,\qquad
\{X_\lambda Y\}=\dd Y+3\lambda Y,\qquad
\{Y_\lambda Y\}=0 .
\end{equation}
Thus $R$ is the semidirect product of the Virasoro conformal algebra $\C[\dd]X$ of central charge zero with its module $\C[\dd]Y$ of conformal weight $3$. Its symmetric algebra
\begin{equation}\label{eq:V}
V=S(R)=\C[X,\dd X,\dd^2X,\ldots]\otimes\Lambda[Y,\dd Y,\dd^2Y,\ldots]
\end{equation}
is a Poisson vertex algebra, with the $\lambda$-bracket extended from $R$ by the Leibniz rules \cite{BDK09}. The algebra $A$ is the quotient of $V$ by the smallest Poisson vertex ideal $I$ containing $X^2$. A short computation with the $\lambda$-brackets shows that $I$ is the differential ideal generated by $X^2$, $XY$, $X\dd Y$ and $Y\dd Y$, so that
\begin{equation}\label{eq:A-presentation}
A=\frac{\C[X,\dd X,\ldots]\otimes\Lambda[Y,\dd Y,\ldots]}{\langle X^2,\,XY,\,X\dd Y,\,Y\dd Y\rangle_\dd}.
\end{equation}

The algebra $A$ is graded by the ghost number $\gh$, the $B$-number $B$ and the spin $S$ (we reserve $s$ for counting factors of $Y$). The generators have
\begin{equation}\label{eq:gradings-XY}
(\gh,B,S)(X)=(2,1,2),\qquad(\gh,B,S)(Y)=(3,3,3),
\end{equation}
and $\dd$ raises the spin by one. It is convenient to replace the ghost number by $p:=\gh-B$ (not to be confused with \cite[(2.34)]{Kha26}, where $p$ denotes the ghost number), so that
\begin{equation}\label{eq:degrees}
(B,p,S)(\dd^kX)=(1,1,k+2),\qquad (B,p,S)(\dd^kY)=(3,0,k+3).
\end{equation}
The relations are also homogeneous in the number of factors of each kind, so $A=\bigoplus_{r,s\ge0}A_{r,s}$, where $A_{r,s}$ is spanned by the images of monomials with $r$ factors of the form $\dd^kX$ and $s$ factors of the form $\dd^kY$. By~\eqref{eq:degrees}, $A_{r,s}$ lies in $(B,p)=(r+3s,r)$, and $(B,p)$ determines $(r,s)$. We therefore write
\begin{equation}\label{eq:ABp}
A_B^p:=A_{r,s}\ \text{ if }(B,p)=(r+3s,r),\qquad A_B^p:=0\ \text{ if $(B,p)$ is not of this form.}
\end{equation}

\subsection{Relative Lie Algebra Cohomology and the Basic Complex}\label{subsec:basic}
We first recall relative Lie algebra cohomology \cite{CE48,HS53}. Let $\g$ be a Lie algebra, $\mathfrak k\subset\g$ a subalgebra and $M$ a $\g$-module. The relative Chevalley--Eilenberg chains and cochains are
\begin{equation}\label{eq:relCE}
C_*(\g,\mathfrak k;M)=\bigl(\Lambda^\bullet(\g/\mathfrak k)\otimes M\bigr)_{\mathfrak k},\qquad
C^*(\g,\mathfrak k;M)=\Hom_{\mathfrak k}\bigl(\Lambda^\bullet(\g/\mathfrak k),M\bigr).
\end{equation}
The first is a quotient complex of the Chevalley--Eilenberg chain complex $\Lambda^\bullet\g\otimes M$. The second is the subcomplex of $\Hom(\Lambda^\bullet\g,M)$ formed by the $\mathfrak k$-\emph{basic} cochains, those that are horizontal, $\alpha(x,\dots)=0$ for $x\in\mathfrak k$, and $\mathfrak k$-equivariant. We write $H_*(\g,\mathfrak k;M)$ and $H^*(\g,\mathfrak k;M)$ for the (co)homology, and omit $M$ when $M=\C$.\footnote{Conceptually, these are the relative groups $\operatorname{Tor}^{(\g,\mathfrak k)}_*(\C,M)$ and $\Ext^*_{(\g,\mathfrak k)}(\C,M)$ \cite{Hoc56}, and the complexes~\eqref{eq:relCE} follow from the $(\g,\mathfrak k)$-projective resolution $U\g\otimes_{U\mathfrak k}\Lambda^\bullet(\g/\mathfrak k)$ of $\C$.}

For $M=\C$ the absolute cochains form the graded-commutative algebra $\Lambda(\g^*)$, with $\g^*$ in degree one, and the differential is the odd derivation that is the transpose of the bracket on generators:
\begin{equation}\label{eq:CEsign}
(d\xi)(x,y)=\xi([x,y]),\qquad\text{i.e.}\qquad dx^\gamma=\tfrac12f^\gamma_{\alpha\beta}\,x^\alpha x^\beta,
\end{equation}
for a basis $x_\alpha$ of $\g$ with dual basis $x^\alpha$ and $[x_\alpha,x_\beta]=f^\gamma_{\alpha\beta}x_\gamma$. Here $(\xi\wedge\eta)(x,y)=\xi(x)\eta(y)-\xi(y)\eta(x)$. The other common convention, $(d\xi)(x,y)=-\xi([x,y])$, changes $d$ by an overall sign, which affects neither cocycles nor coboundaries. The relative cochains $C^*(\g,\mathfrak k)=\Lambda\bigl((\g/\mathfrak k)^*\bigr)^{\mathfrak k}$ form the subalgebra of basic elements. If $\g=\mathfrak k\oplus\mathfrak p$ for a $\mathfrak k$-stable complement $\mathfrak p$, then $(\g/\mathfrak k)^*\cong\mathfrak p^*$. Since a horizontal cochain only sees the $\mathfrak p$-components of brackets, $d$ acts on $C^*(\g,\mathfrak k)=\Lambda(\mathfrak p^*)^{\mathfrak k}$ through the derivation of $\Lambda(\mathfrak p^*)$ given by
\begin{equation}\label{eq:relative-d}
(d\xi)(x,y)=\xi\bigl([x,y]_{\mathfrak p}\bigr),\qquad x,y\in\mathfrak p .
\end{equation}
In general this derivation squares to zero only on $\mathfrak k$-invariants. If $\mathfrak p$ is an ideal, it is the Chevalley--Eilenberg differential of $\mathfrak p$, and restriction along $\mathfrak p\hookrightarrow\g$ gives an isomorphism of complexes
\begin{equation}\label{eq:relideal}
C^*(\g,\mathfrak k)\cong C^*(\mathfrak p)^{\mathfrak k}.
\end{equation}

For the infinite-dimensional Lie algebras of this section we use cochains that are polynomial in the coordinates: if $\g=\bigoplus_{n\le0}\g_n$ is graded with finite-dimensional components and $\mathfrak k$ is a graded subalgebra, we replace $\g^*$ by the graded dual $\g^\vee=\bigoplus_n\g_n^*$. The transpose of the bracket maps $\g_n^*$ into the finite sum $\bigoplus_{i+j=n}(\g_i\otimes\g_j)^*$, so $\Lambda(\g^\vee)$ is a subcomplex of $\Lambda(\g^*)$, and we define $C^*(\g,\mathfrak k)$ as its subcomplex of basic elements. Graded modules $M$ are treated in the same way. For finite-dimensional $M$ see the beginning of \S\ref{sec:inputs}.

We now specialize to the Lie algebra of the introduction: $R=\C[z,w]/(w^2)$, $\g=\sltwo\otimes R=\sltwo[z,w]/(w^2)$ and $\mathfrak k=\sltwo\otimes1$. We identify $\sltwo\cong\mathfrak{so}_3$, with basis $e_1,e_2,e_3$, $[e_a,e_b]=\epsilon_{abc}e_c$ and invariant form $\delta_{ab}$. Since the invariant form is $\delta_{ab}$, the position of a color index carries no information: repeated indices are summed, and we place indices so that free indices match. Then $\g$ has the basis of \emph{modes} $e_az^n$ and $we_az^j$, $n,j\ge0$, and we grade it by $\deg z=\deg w=-1$, as in the introduction. Let $\mathfrak m=(z,w)\subset R$ be the maximal ideal of the origin. Since $R=\C\oplus\mathfrak m$, the constant subalgebra has the complementary ideal $\sltwo\otimes\mathfrak m$, and
\begin{equation}\label{eq:g-decomp}
\g=\sltwo\ltimes(\sltwo\otimes\mathfrak m),\qquad \sltwo\otimes\mathfrak m=z\,\sltwo[z]\ltimes w\,\sltwo[z],
\end{equation}
where $w\,\sltwo[z]$ is an abelian ideal because $w^2=0$. Let $c_n^a$ ($n\ge1$) and $b_j^a$ ($j\ge0$) be the coordinates on $\sltwo\otimes\mathfrak m$ dual to the modes $e_az^n$ and $we_az^j$. By~\eqref{eq:relideal}, the relative cochains form the \emph{basic complex}
\begin{equation}\label{eq:C}
\cC:=C^*(\g,\sltwo)=C^*(\sltwo\otimes\mathfrak m)^{\sltwo}=\Lambda\bigl(b_j^a,\,c_n^a:\ j\ge0,\ n\ge1,\ a=1,2,3\bigr)^{\sltwo},
\end{equation}
the invariant part of the exterior algebra on the odd generators $b_j^a$ and $c_n^a$. We denote its differential by $Q$. The brackets of modes are
\begin{equation}\label{eq:mode-brackets}
[e_bz^i,e_cz^k]=\epsilon_{bcd}\,e_dz^{i+k},\qquad [e_bz^i,we_cz^k]=\epsilon_{bcd}\,we_dz^{i+k},\qquad [we_bz^i,we_cz^k]=0,
\end{equation}
so by~\eqref{eq:CEsign} the differential is given on generators by
\begin{equation}\label{eq:Q}
Qc_n^a=\frac12\sum_{i=1}^{n-1}\epsilon^{abc}\,c_i^bc_{n-i}^c,\qquad
Qb_j^a=\sum_{i=1}^{j}\epsilon^{abc}\,c_i^bb_{j-i}^c,
\end{equation}
and these formulas define a differential on the whole exterior algebra, the Chevalley--Eilenberg differential of $\sltwo\otimes\mathfrak m$.

The cochain degree is the ghost number $\gh=B+p$, where $B=\#b$ and $p=\#c$. The spin $S$ is the grading induced by $\deg z=\deg w=-1$, so that $S(b_j^a)=j+1$ and $S(c_n^a)=n$, and the $B$-number is the $w$-degree, induced by $\deg z=0$, $\deg w=-1$. The differential has degree $(0,1,0)$ in $(B,p,S)$, and each component of fixed $(B,p,S)$ is finite-dimensional. In particular, $\cC$ is the graded dual of the relative chain complex $C_*(\g,\sltwo)$. We put
\begin{equation}\label{eq:H}
H_B^p:=H^p(\cC_B,Q),\qquad H:=\bigoplus_{B,p}H_B^p=\Obs(\sltwo),
\end{equation}
so that, in the sense above,
\begin{equation}\label{eq:Hdef}
H^*\bigl(\sltwo[z,w]/(w^2),\sltwo\bigr)=H^*(\cC,Q)=H .
\end{equation}
Since $Q$ preserves the spin and each spin component is finite-dimensional, the complex of all relative cochains, polynomial or not, is the product of the spin components of $\cC$, so restricting to polynomial cochains only replaces a product of cohomology groups by their direct sum.

Because $w\,\sltwo[z]$ is an abelian ideal of $\g=\sltwo[z]\ltimes w\,\sltwo[z]$, the transpose of the bracket maps coordinates on $\sltwo[z]$ to quadratic expressions in themselves, and coordinates on $w\,\sltwo[z]$ to products of a coordinate on $\sltwo[z]$ with one on $w\,\sltwo[z]$. Thus $Q$ preserves $B$, and $\cC_B$ is the relative complex of $\sltwo[z]$ with coefficients in the $B$-th exterior power of the coadjoint module $(w\,\sltwo[z])^\vee$:
\begin{equation}\label{eq:CB-coeff}
\cC_B=C^*\bigl(\sltwo[z],\sltwo;\Lambda^B(w\,\sltwo[z])^\vee\bigr),\qquad
H_B^p=H^p\bigl(\sltwo[z],\sltwo;\Lambda^B(w\,\sltwo[z])^\vee\bigr).
\end{equation}
We use the same notation $H^*(\sltwo\otimes R,\sltwo)$ in \S\ref{sec:deformation} for $R=\C[z,w]/(w^2-z)$ and $R=\C[z,w]/(w^2-1)$, which have the same basis of modes $z^n$, $wz^j$.

In the language of \cite[\S3]{Kha26}, whose conventions we follow, $\cC$ is the basic part of a classical $bc$-system and $Q$ is its BRST differential. The generators of $\cC$ are the Taylor coefficients
\begin{equation}\label{eq:modes}
b_j^a=\tfrac1{j!}\dd^jb^a\quad(j\ge0),\qquad c_n^a=\tfrac1{n!}\dd^nc^a\quad(n\ge1),
\end{equation}
of the ghost fields $b^a$ and $c^a$, of spins $1$ and $0$. The zero mode $c^a$ itself does not occur, because cochains are horizontal for the constant subalgebra.

Two families of operators on $\cC$ will be used throughout, and both come from derivations of $\g$. In general, a homogeneous derivation $D$ of $\g$ with $D(\mathfrak k)\subset\mathfrak k$ acts on $C^*(\g,\mathfrak k)$ by the even derivation $\theta_D$ with $\theta_D\xi=\xi\circ D$ on generators: $\theta_D$ preserves basic cochains, and it commutes with the differential because the latter is the transpose of the bracket. Both of the derivations below vanish on $\mathfrak k=\sltwo$. First, the derivation $\dd_z$ of $R$ with $\dd_zw=0$ gives the even derivation
\begin{equation}\label{eq:translation}
\dd b_j^a=(j+1)\,b_{j+1}^a,\qquad \dd c_n^a=(n+1)\,c_{n+1}^a
\end{equation}
of $\cC$. It commutes with $Q$ and acts on $H$. It is the translation of the Poisson vertex algebra $\Obs(\sltwo)$. Since $\dd_z$ does not preserve $\sltwo\otimes\mathfrak m$, the derivations~\eqref{eq:Q} and~\eqref{eq:translation} commute only on invariants: on the whole exterior algebra one finds
\begin{equation}\label{eq:Q-translation}
[Q,\dd]=-c_1^aL_a,
\end{equation}
where $L_a$ is the derivation by which the constant mode $e_a$ acts on the coordinates, $L_ab_j^c=\epsilon_a{}^c{}_d\,b_j^d$ and $L_ac_n^c=\epsilon_a{}^c{}_d\,c_n^d$. Second, for $m\ge1$, the endomorphism of $\g$ that is multiplication by $z^m$ on $w\,\sltwo[z]$ and zero on $\sltwo[z]$ is a derivation, because $w\,\sltwo[z]$ is an abelian ideal and $[xf,wyg]=w[x,y]fg$ is $\C[z]$-bilinear. It gives the even derivations
\begin{equation}\label{eq:T}
T_mb_j^a=b_{j-m}^a\quad(\text{zero if }j<m),\qquad T_mc_n^a=0,
\end{equation}
which preserve $\sltwo\otimes\mathfrak m$ and therefore commute with $Q$ on the whole exterior algebra. The derivation $T_m$ lowers the mode of each $b$-ghost by $m$. It is the transpose of multiplication by $z^m$ on the coefficient module $w\,\sltwo[z]$. In \S\ref{subsec:positions} it will measure the positions of the $b$-ghosts.

\subsection{The Map \texorpdfstring{$\varphi$}{phi}}\label{subsec:phi}
Put
\begin{equation}\label{eq:XY}
\Xc:=-\delta_{ab}\,b_0^ac_1^b,\qquad \Yc:=b_0^1b_0^2b_0^3=\tfrac16\,\epsilon_{abc}\,b_0^ab_0^bb_0^c .
\end{equation}
Since rescaling $Y$ by a nonzero constant is an automorphism of $A$ (the relations and the $\lambda$-brackets are homogeneous in $Y$), the normalization of $\Yc$, which differs from that of \cite{Kha26} by a factor $1/6$, is immaterial for Theorem~\ref{thm:main}.

\begin{lemma}\label{lem:relations}
The cochains $\Xc$ and $\Yc$ are $Q$-closed, and
\begin{equation}\label{eq:relations}
\Xc^2=Q\bigl(-\epsilon_{abc}\,b_0^ab_0^bc_2^c\bigr),\qquad \Xc\Yc=0,\qquad
\Xc\,\dd\Yc=Q\Bigl(-\tfrac14\bigl(\delta_{ab}\,b_0^ab_1^b\bigr)^2\Bigr),\qquad \Yc\,\dd\Yc=0 .
\end{equation}
\end{lemma}

\begin{proof}
By~\eqref{eq:Q}, $Qb_0^a=Qc_1^a=0$. Every term of $\Xc\Yc$ and of $\Yc\,\dd\Yc$ contains a repeated $b_0^a$. Using $Qc_2^c=\frac12\epsilon^{cde}c_1^dc_1^e$ and $\epsilon_{abc}\epsilon_{cde}=\delta_{ad}\delta_{be}-\delta_{ae}\delta_{bd}$ we find
\begin{equation}\label{eq:X2-exact}
Q\bigl(\epsilon_{abc}b_0^ab_0^bc_2^c\bigr)=b_0^ab_0^bc_1^ac_1^b=-\Xc^2 .
\end{equation}
Similarly, $\sigma:=\delta_{ab}\,b_0^ab_1^b$ satisfies $Q\sigma=-\epsilon_{abc}b_0^ab_0^bc_1^c$, hence
\begin{equation}\label{eq:XdY-exact}
Q(\sigma^2)=2\sigma\,Q\sigma=-4\,\Yc\,\delta_{ab}\,b_1^ac_1^b=-4\,\Xc\,\dd\Yc .\qedhere
\end{equation}
\end{proof}

Consequently, the assignment $X\mapsto[\Xc]$, $Y\mapsto[\Yc]$, extended multiplicatively and $\dd$-equivariantly, defines a homomorphism of differential algebras
\begin{equation}\label{eq:phi}
\varphi:A\longrightarrow H .
\end{equation}
Up to the normalization of $Y$, this is the map of \cite[\S3.6]{Kha26}. In particular it intertwines the $\lambda$-brackets of $A$ with the Poisson vertex algebra structure of $\Obs(\sltwo)$, which is induced from the $bc$-system with $\{b^a{}_\lambda c^b\}=\delta^{ab}$. Indeed, writing $b^a=b_0^a$ and $\dd c^a=c_1^a$, one finds
\begin{equation}\label{eq:X-ghosts}
\{\Xc_\lambda b^a\}=(\dd+\lambda)b^a,\qquad\{\Xc_\lambda\dd c^a\}=(\dd+\lambda)\dd c^a,
\end{equation}
and the Leibniz rule gives
\begin{equation}\label{eq:XY-brackets-cochains}
\{\Xc_\lambda\Xc\}=(\dd+2\lambda)\Xc,\qquad\{\Xc_\lambda\Yc\}=(\dd+3\lambda)\Yc,\qquad\{\Yc_\lambda\Yc\}=0.
\end{equation}
Thus the map from $V$ preserves $\lambda$-brackets and kills $X^2$, so its kernel is a Poisson vertex ideal containing $X^2$, and therefore contains $I$. Lemma~\ref{lem:relations} makes the vanishing on the generators of $I$ explicit. The map $\varphi$ preserves the gradings~\eqref{eq:degrees}, so Theorem~\ref{thm:main} is equivalent to the bijectivity of
\begin{equation}\label{eq:phi-Bp}
\varphi:A_B^p\longrightarrow H_B^p\qquad\text{for all }(B,p).
\end{equation}

\subsection{Dual Formulation}\label{subsec:dual}
With graded duals as in \S\ref{subsec:conventions}, put
\begin{equation}\label{eq:psi}
K_{B,p}:=(H_B^p)^\vee,\qquad A^\vee_{B,p}:=(A_B^p)^\vee,\qquad
\psi_{B,p}:=\varphi^\vee:\ K_{B,p}\longrightarrow A^\vee_{B,p}.
\end{equation}
Theorem~\ref{thm:main} is equivalent to the statement that every $\psi_{B,p}$ is an isomorphism. We show that every $\psi_{B,p}$ is surjective in \S\ref{sec:injective}, and that it is injective by induction on $B$ in \S\S\ref{sec:sheaves}--\ref{sec:proof}.

\section{Representation-Theoretic Input}\label{sec:inputs}

Before turning to the proof, let us collect the results from the representation theory of current algebras that we will need. Throughout, $\lie:=\sltwo[z]$, $V_m$ is the irreducible $\sltwo$-module of highest weight $m$ (of dimension $m+1$), and for $a\in\C$, $V_m(a)$ is the evaluation module of $\lie$ on which $x\,f(z)$ acts by $f(a)\,x$. For an $\lie$-module $M$ we write
\begin{equation}\label{eq:rel-chains-l}
C_*(\lie,\sltwo;M)=\bigl(\Lambda^\bullet(\lie/\sltwo)\otimes M\bigr)_{\sltwo},\qquad
C^*(\lie,\sltwo;M)=\Hom_{\sltwo}\bigl(\Lambda^\bullet(\lie/\sltwo),M\bigr)
\end{equation}
for the relative chains and cochains of~\eqref{eq:relCE}. When $M$ is finite-dimensional we use the full $\Hom$, that is, ordinary algebraic cochains. When $M$ is graded with finite-dimensional graded pieces, as in \S\ref{subsec:basic}, we use graded duals.

\begin{theorem}[{Relative BGG resolution \cite{GL76,Lep79}}]\label{thm:BGG}
The trivial $\lie$-module has a resolution by $(\lie,\sltwo)$-relatively projective modules
\begin{equation}\label{eq:BGG-resolution}
\cdots\to P_2\to P_1\to P_0\to\C\to0,\qquad
P_p=U(\lie)\otimes_{U(\sltwo)}V_{2p}[d_p],\qquad d_p=\tfrac{p(p+1)}2,
\end{equation}
where $[d_p]$ shifts the polynomial-mode degree by $d_p$. Consequently, if the constant subalgebra $\sltwo$ acts locally finitely on an $\lie$-module $M$, then $H_*(\lie,\sltwo;M)$ and $H^*(\lie,\sltwo;M)$ are computed by complexes with terms
\begin{equation}\label{eq:BGG-terms}
(M\otimes V_{2p})^{\sltwo}[d_p]\qquad\text{and}\qquad\Hom_{\sltwo}(V_{2p},M)[-d_p],\qquad p\ge0,
\end{equation}
respectively.
\end{theorem}

The underlying computation,
\begin{equation}\label{eq:GL}
H_p\bigl(z\,\sltwo[z]\bigr)\cong V_{2p}[d_p],
\end{equation}
is the affine $A_1$ case of the theorem of Garland and Lepowsky \cite{GL76}, the affine analogue of Kostant's theorem \cite{Kos61}. The resolution by generalized Verma modules, together with the resulting complex computing relative (co)homology, is due to Lepowsky \cite[Proposition~5.3, Corollary~5.4 and Theorem~5.5]{Lep79}. We use it after exchanging positive and negative currents, which reverses the sign of the mode grading. See also \cite{Kum02}. Since $V_{2p}^{\sltwo}=0$ for $p>0$, trivial coefficients give
\begin{equation}\label{eq:trivial}
H^*(\sltwo[z],\sltwo;\C)=\C,
\end{equation}
concentrated in cohomological degree and polynomial-mode degree zero.

\begin{lemma}[{Hochschild--Serre, cf.\ \cite{HS53}}]\label{lem:HS}
Let $\mathfrak a=\mathfrak k\ltimes\mathfrak n$ with $\mathfrak k$ semisimple, and let $M$ be an $\mathfrak a$-module such that each cochain space $C^q(\mathfrak a;M)$ is a locally finite $\mathfrak k$-module. Then
\begin{equation}\label{eq:HS}
H^*(\mathfrak a;M)\cong H^*(\mathfrak k;\C)\otimes H^*(\mathfrak a,\mathfrak k;M).
\end{equation}
\end{lemma}

This is proved as in \cite{HS53}: one decomposes the cochains into $\mathfrak k$-isotypic components. The nontrivial components are contracted using the Casimir operator, and the trivial component contributes $H^*(\mathfrak k;\C)$. The hypothesis holds in all cases below. For full-$\Hom$ cochains of $\lie$ with finite-dimensional coefficients $M$, the space $C^q$ is a product of finite-dimensional $\sltwo$-modules with highest weights at most $2q$ plus the largest weight of $M$. Only finitely many isotypic types occur, although with infinite multiplicities. For graded cochains the hypothesis holds degree by degree.

\begin{theorem}[{Teleman \cite[Theorem~0]{Tel95}, \cite[Theorem~3.1]{Tel96}}]\label{thm:eval}
Let $a_1,\dots,a_n\in\C$ be pairwise distinct and $m_1,\dots,m_n\ge0$. For full-$\Hom$ relative cochains,
\begin{equation}\label{eq:Teleman}
H^j\Bigl(\lie,\sltwo;\bigotimes_{i=1}^nV_{m_i}(a_i)\Bigr)\cong
\begin{cases}\C,& \text{if } m_i=2p_i\text{ for all }i\text{ and } j=\sum_ip_i,\\[2pt] 0,&\text{otherwise.}\end{cases}
\end{equation}
\end{theorem}

This is the case $\Sigma=\C$, $h=0$ of \cite[Theorem~3.1]{Tel96}, the genus-zero statement recalled there from \cite[\S1.4, Theorem~0]{Tel95}. That theorem computes the cohomology of the pair $(\g[z],\g)$ with coefficients in $\mathsf H\otimes\bigotimes_kV_k(z_k)$, where the points $z_k$ are distinct, the $V_k$ are irreducible, and $\mathsf H$ is a positive-energy representation at $\infty$ of any level $h\ge0$ \cite[\S1]{Tel96}. The cohomology is that of the standard complex $\Hom_\g(\Lambda^\bullet(\g[z]/\g),-)$ \cite[(3.1.4)]{Tel95}, which is our full-$\Hom$ complex. At level zero $\mathsf H$ is trivial. For $\sltwo$ the dual Coxeter number is $c=2$, and the $\rho$-shifted highest weight of $V_m$ pairs with the positive root to give $m+1$. The alcove walls are where this pairing lies in $(h+c)\mathbb Z=2\mathbb Z$. Thus $V_m$ is singular if $m$ is odd, and then the cohomology vanishes. If $m=2p$, the shifted weight lies in the alcove $(2p,2p+2)$, which is separated from the positive alcove $(0,2)$ by the $p$ walls $2,4,\dots,2p$. So $V_{2p}$ is regular of length $p$, and the small representation obtained from it by reflection \cite[\S2]{Tel96} is $\C$, the only small representation at level zero. By \cite[Theorem~3.1]{Tel96} the cohomology is then concentrated in degree $\sum_ip_i$, where it is $\C^{\lie}=\C$. We do not use any multiplicative property of Theorem~\ref{thm:eval}: the factorization of cup products that we need is proved in Lemma~\ref{lem:cup}.

\section{Injectivity: Separating Cycles}\label{sec:injective}

Let us now turn to the proof of Theorem~\ref{thm:main}, starting with the easier half: injectivity of $\varphi$, or equivalently surjectivity of every $\psi_{B,p}$. For this we first need a concrete description of the graded dual of $A$.

\subsection{The Dual of \texorpdfstring{$A$}{A}}
We describe the graded dual of $A$ by means of the generating fields
\begin{equation}\label{eq:XY-fields}
X(z):=\sum_{k\ge0}\dd^kX\,\frac{z^k}{k!},\qquad Y(z):=\sum_{k\ge0}\dd^kY\,\frac{z^k}{k!}.
\end{equation}
Let $V_{r,s}\subset V$ be the span of monomials with $r$ factors $\dd^kX$ and $s$ factors $\dd^kY$. For a functional $\eta$ on $V_{r,s}$ put
\begin{equation}\label{eq:Omega}
\Omega_\eta(z_1,\dots,z_r;y_1,\dots,y_s):=\eta\bigl(X(z_1)\cdots X(z_r)\,Y(y_1)\cdots Y(y_s)\bigr).
\end{equation}
If $\eta$ belongs to the graded dual, $\Omega_\eta$ is a polynomial, symmetric in the $z_i$ and alternating in the $y_a$, and $\eta\mapsto\Omega_\eta$ identifies the graded dual of $V_{r,s}$ with the space of such polynomials. A monomial of degree $d$ pairs with elements of spin $2r+3s+d$. Put
\begin{equation}\label{eq:F}
\cO_{r,s}:=\C[z_1,\dots,z_r]^{S_r}\otimes\C[y_1,\dots,y_s]^{S_s},\qquad
F_{r,s}:=\Delta(z)^2\,\Delta(y)^3\prod_{i=1}^r\prod_{a=1}^s(z_i-y_a)^2,
\end{equation}
where $\Delta(z)=\prod_{i<j}(z_i-z_j)$, and empty products are equal to one. We also write $\C[z,y]$ for the polynomial ring in $z_1,\dots,z_r$ and $y_1,\dots,y_s$.

\begin{lemma}\label{lem:dualA}
The map $\eta\mapsto\Omega_\eta$ identifies $A_{r,s}^\vee$ with $F_{r,s}\,\cO_{r,s}$.
\end{lemma}

\begin{proof}
This is a reformulation of \cite[(2.54)]{Kha26}. We include the short argument. A functional on $V_{r,s}$ descends to $A_{r,s}$ if and only if it annihilates $I\cap V_{r,s}$, which is spanned by products of derivatives of $X^2$, $XY$, $X\dd Y$, $Y\dd Y$ with monomials. Since
\begin{equation}\label{eq:exp-relations}
X(z)^2=e^{z\dd}(X^2),\quad X(z)Y(z)=e^{z\dd}(XY),\quad X(z)Y'(z)=e^{z\dd}(X\dd Y),\quad Y(z)Y'(z)=e^{z\dd}(Y\dd Y),
\end{equation}
this happens if and only if, for all $i\ne j$ and $a\ne b$,
\begin{enumerate}[label=(\alph*),leftmargin=2em]
\item $\Omega_\eta$ vanishes on $z_i=z_j$;
\item $\Omega_\eta$ and $\dd_{y_a}\Omega_\eta$ vanish on $y_a=z_i$;
\item $\dd_{y_b}\Omega_\eta$ vanishes on $y_a=y_b$.
\end{enumerate}
By symmetry, (a) is equivalent to divisibility by $(z_i-z_j)^2$, and (b) is equivalent to divisibility by $(z_i-y_a)^2$. Since $\Omega_\eta$ is alternating, it equals $(y_a-y_b)g$ with $g$ symmetric in $y_a,y_b$, and (c) says that $g$ vanishes on $y_a=y_b$. Thus (c) is equivalent to divisibility by $(y_a-y_b)^3$. The linear forms $z_i-z_j$, $z_i-y_a$, $y_a-y_b$ are pairwise non-associate primes, so (a)--(c) together say that $F_{r,s}$ divides $\Omega_\eta$, and the quotient is symmetric in the $z$'s and in the $y$'s.
\end{proof}

\subsection{Cartan Currents}
Choose a Cartan element $h$ and root vectors $E,F$ of $\sltwo$, normalized here by $[h,E]=E$ and $[h,F]=-F$, let $\hh=\C h$, and consider the subalgebra and pair
\begin{equation}\label{eq:Cartan-pair}
\tl:=\hh[z]\ltimes w\,\sltwo[z]\ \subset\ \sltwo[z,w]/(w^2),\qquad (\tl,\hh)\ \subset\ \bigl(\sltwo[z,w]/(w^2),\sltwo\bigr).
\end{equation}
Restriction of basic cochains commutes with the differentials. Dually, the inclusion of relative chains $\iota_*:C_*(\tl,\hh)\to C_*(\sltwo[z,w]/(w^2),\sltwo)$ is a chain map. Since $\cC$ is the graded dual of the second complex, every cycle $\gamma\in C_*(\tl,\hh)$ defines a functional
\begin{equation}\label{eq:cycle-functional}
H\longrightarrow\C,\qquad [\alpha]\longmapsto\langle\alpha,\iota_*\gamma\rangle=\langle\iota^*\alpha,\gamma\rangle,
\end{equation}
which is well defined because coboundaries pair to zero with cycles.

The relative chains are the weight-zero part of an exterior algebra,
\begin{equation}\label{eq:Cartan-chains}
C_*(\tl,\hh)=\Lambda^\bullet\bigl(z\hh[z]\oplus w\,\sltwo[z]\bigr)_{\hh}.
\end{equation} Write $\xi_k:=hz^k$ for $k\ge1$. The only brackets in $\tl$ between the generators of this exterior algebra which do not vanish are
\begin{equation}\label{eq:Cartan-brackets}
[\xi_k,wEz^m]=wEz^{m+k},\qquad [\xi_k,wFz^m]=-wFz^{m+k}.
\end{equation}
Thus the chain complex is a tensor product of three pieces: the exterior algebra on the $\xi_k$, the \emph{root part} $\Lambda(wE\otimes\C[z])\otimes\Lambda(wF\otimes\C[z])$, and the \emph{neutral part} $\Lambda(wh\otimes\C[z])$, which is inert. Weight zero forces equally many factors $s$ of each root color. Identifying wedges with alternants,
\begin{equation}\label{eq:alternant}
wEz^{m_1}\wedge\cdots\wedge wEz^{m_s}\longmapsto\det\bigl(u_a^{m_b}\bigr),
\end{equation}
and similarly for $F$ with variables $v_1,\dots,v_s$, the root part with $s$ factors of each color becomes $\Delta(u)\Delta(v)\,\cO^\Car_s$, where
\begin{equation}\label{eq:OCar}
\cO^\Car_s:=\C[u_1,\dots,u_s]^{S_s}\otimes\C[v_1,\dots,v_s]^{S_s}.
\end{equation}
Under this identification $\xi_k$ acts by multiplication by
\begin{equation}\label{eq:fk}
f_k:=p_k(u)-p_k(v),\qquad p_k(u)=\textstyle\sum_au_a^k .
\end{equation}
Therefore, up to signs, the chain complex is the Koszul complex
\begin{equation}\label{eq:koszul}
\cO^\Car_s\otimes\Lambda(\xi_1,\xi_2,\ldots),\qquad d\xi_k=f_k,
\end{equation}
tensored with the neutral part, on which the differential vanishes.

We now produce cycles by passing from power sums to elementary symmetric functions. Introduce a formal parameter $\tau$ and put
\begin{gather}
\label{eq:E-uv}\mathcal E_u(\tau)=\prod_a(1-u_a\tau),\qquad \mathcal E_v(\tau)=\prod_a(1-v_a\tau),\\
\label{eq:L-Xi}L(\tau):=\log\frac{\mathcal E_v(\tau)}{\mathcal E_u(\tau)}=\sum_{k\ge1}\frac{f_k}{k}\,\tau^k,\qquad
\Xi(\tau):=\sum_{k\ge1}\frac{\xi_k}{k}\,\tau^k,\\
\label{eq:Theta}\Theta(\tau):=\mathcal E_v(\tau)\,\frac{e^{-L(\tau)}-1}{L(\tau)}\;\Xi(\tau)=\sum_{k\ge1}\theta_k\tau^k,
\end{gather}
where $(e^{-L}-1)/L$ stands for $\sum_{n\ge0}(-1)^{n+1}L^n/(n+1)!$. Since $L$ has no constant term, each $\theta_k$ is a finite expression
\begin{equation}\label{eq:theta-k}
\theta_k=-\frac{\xi_k}{k}+\sum_{j<k}\pi_{kj}\,\xi_j,\qquad \pi_{kj}\in\cO^\Car_s .
\end{equation} The Koszul differential is $\cO^\Car_s$-linear with $d\Xi=L$, so
\begin{equation}\label{eq:dtheta}
d\Theta=\mathcal E_v\bigl(e^{-L}-1\bigr)=\mathcal E_u-\mathcal E_v,\qquad\text{i.e.}\qquad
d\theta_k=\begin{cases}(-1)^k\bigl(e_k(u)-e_k(v)\bigr),&k\le s,\\ 0,&k>s.\end{cases}
\end{equation}
In particular $\theta_{s+1},\theta_{s+2},\dots$ are cycles. (For $s=0$ one simply has $\theta_k=-\xi_k/k$.)

To pair these cycles with $\varphi(A)$, we use the generating functions
\begin{equation}\label{eq:beta-zeta}
\beta_x(z)=\sum_{j\ge0}(wxz^j)^\vee z^j\quad(x\in\{h,E,F\}),\qquad \zeta_h(z)=\sum_{n\ge1}(hz^n)^\vee z^n
\end{equation}
of the coordinates on $\tl/\hh$ dual to the basis $\{hz^n,\,wxz^j\}$. With $b^a(z)=\sum_jb_j^az^j$ and $c^a(z)=\sum_{n\ge1}c_n^az^n$, the fields of $\Xc$ and $\Yc$ are
\begin{equation}\label{eq:XY-cochain-fields}
\Xc(z):=\sum_k\dd^k\Xc\,\frac{z^k}{k!}=-\delta_{ab}\,b^a(z)\,\frac{d}{dz}c^b(z),\qquad \Yc(z)=b^1(z)b^2(z)b^3(z).
\end{equation}
Restricting them to $\tl$ kills the root components of $c$, and gives
\begin{equation}\label{eq:restrictedfields}
\iota^*\Xc(z)=\kappa_X\,\beta_h(z)\,\zeta_h'(z),\qquad \iota^*\Yc(y)=\kappa_Y\,\beta_h(y)\beta_E(y)\beta_F(y)
\end{equation}
for nonzero constants $\kappa_X,\kappa_Y$. For $0\le n_1<\dots<n_r$ and $G\in\C[u_1,\dots,u_s]^{S_s}$ consider the chain
\begin{equation}\label{eq:gamma-nG}
\gamma_{n,G}:=\theta_{s+1+n_1}\wedge\cdots\wedge\theta_{s+1+n_r}\wedge\bigl(\Delta(u)\Delta(v)\,G\bigr)\wedge\varpi,\qquad
\varpi:=wh\wedge whz\wedge\cdots\wedge whz^{r+s-1}.
\end{equation}
It is a cycle, because the Koszul differential is an $\cO^\Car_s$-linear derivation of~\eqref{eq:koszul} and each $\theta_{s+1+n_j}$ is closed by~\eqref{eq:dtheta}. Note that closedness holds identically in $u,v$. The specialization $u=v$ below is only used to evaluate pairings.

\begin{lemma}\label{lem:pairing}
For a nonzero constant $\kappa$ independent of $n$ and $G$,
\begin{equation}\label{eq:pairing}
\bigl\langle\varphi\bigl(X(z_1)\cdots X(z_r)\,Y(y_1)\cdots Y(y_s)\bigr),\ \iota_*\gamma_{n,G}\bigr\rangle
=\pm\kappa\;F_{r,s}\;s_n(z)\;G(y),
\qquad s_n(z):=\frac{\det(z_i^{n_j})}{\Delta(z)} .
\end{equation}
\end{lemma}

\begin{proof}
By~\eqref{eq:restrictedfields} the restricted cochain is a product of coordinate functions of four kinds, $\zeta_h'$, $\beta_h$, $\beta_E$ and $\beta_F$, which pair only with chain factors of the corresponding kinds $\xi_k$, $whz^j$, $wEz^j$ and $wFz^j$. The pairing therefore factors into three determinants. (i) The neutral factor gives a Vandermonde determinant in all $r+s$ variables,
\begin{equation}\label{eq:neutral-pairing}
\Bigl\langle\prod_i\beta_h(z_i)\prod_a\beta_h(y_a),\,\varpi\Bigr\rangle=\pm\Delta(z_1,\dots,z_r,y_1,\dots,y_s).
\end{equation} (ii) Pairing $\prod_a\beta_E(y_a)$ with an alternant $\rho(u)$ returns $\pm\rho(y)$, and similarly for $F$, so a root element $\Delta(u)\Delta(v)P(u,v)$ pairs to $\pm\Delta(y)^2\,\ev(P)$, where $\ev:\cO^\Car_s\to\C[y]^{S_s}$ is the specialization $u=v=y$. (iii) The coefficients $\pi_{kj}\in\cO^\Car_s$ of the $\theta$'s multiply the root factor, so by (ii) they enter through $\ev$, which is a ring homomorphism. Since
\begin{equation}\label{eq:Xi-pairing}
\langle\Xi(\tau),\zeta_h'(z)\rangle=\sum_k\frac{\tau^k}{k}\,kz^{k-1}=\frac{\tau}{1-z\tau},
\end{equation}
and the prefactor of $\Xi$ in $\Theta$ specializes to $-\mathcal E_y(\tau)$ at $u=v=y$,
\begin{equation}\label{eq:cyclepairing}
\ev\langle\theta_{s+1+n},\zeta_h'(z)\rangle=-[\tau^{s+n}]\,\frac{\mathcal E_y(\tau)}{1-z\tau}=-z^n\prod_{a=1}^s(z-y_a).
\end{equation}
The Cartan factor is therefore
\begin{equation}\label{eq:Cartan-factor}
\det\bigl(\ev\langle\theta_{s+1+n_j},\zeta_h'(z_i)\rangle\bigr)=\pm\prod_{i,a}(z_i-y_a)\,\det(z_i^{n_j}).
\end{equation} Multiplying (i), (ii) and (iii), and using $\Delta(z,y)=\Delta(z)\Delta(y)\prod_{i,a}(z_i-y_a)$, gives $\pm F_{r,s}\,s_n(z)\,G(y)$ up to the product of the constants $\kappa_X^r\kappa_Y^s$ and combinatorial factors.
\end{proof}

\begin{proposition}\label{prop:injective}
The map $\varphi$ is injective. Equivalently, every $\psi_{B,p}$ is surjective.
\end{proposition}

\begin{proof}
As $n$ ranges over strictly increasing sequences, the Schur polynomials $s_n(z)$ form a basis of $\C[z_1,\dots,z_r]^{S_r}$ \cite[I.3]{Mac95}, and $G(y)$ ranges over $\C[y_1,\dots,y_s]^{S_s}$. By Lemma~\ref{lem:pairing}, the functionals $a\mapsto\langle\varphi(a),\iota_*\gamma_{n,G}\rangle$ on $A_{r,s}$ correspond to polynomials spanning $F_{r,s}\cO_{r,s}$, which is all of $A_{r,s}^\vee$ by Lemma~\ref{lem:dualA}. Therefore an element $a\in A_{r,s}$ with $\varphi(a)=0$ pairs to zero with all of $A^\vee_{r,s}$, so $a=0$. Different $(r,s)$ lie in different $(B,p)$, and $\varphi$ preserves $(B,p)$.
\end{proof}

\section{Sheaves on Configuration Spaces of $b$-Ghosts}\label{sec:sheaves}

In the previous section we showed that every $\psi_{B,p}$ is surjective. Sections~\ref{sec:sheaves}--\ref{sec:proof} show that it is also injective. As explained in \S\ref{subsec:idea}, the idea is to regard both sides as sheaves on the configuration space of $b$-ghosts, and we now set this up.

\subsection{The Finite BGG Complex}\label{subsec:BGG}
As an $\lie$-module, $w\,\sltwo[z]\cong\sltwo\otimes\C[z]$, with
\begin{equation}\label{eq:a-action}
(xz^m)\cdot(wyz^n)=w[x,y]z^{m+n}.
\end{equation}
Put
\begin{equation}\label{eq:NB}
N_B:=\Lambda^B\bigl(w\,\sltwo[z]\bigr),\qquad S(wyz^j)=j+1 .
\end{equation}
The $B$-number $B$ part $\cC_B$ of $\cC$ is the graded dual of the relative chain complex $C_*(\lie,\sltwo;N_B)$, and $Q$ is the transpose of its differential. Therefore
\begin{equation}\label{eq:K}
K_{B,p}=H_p(\lie,\sltwo;N_B).
\end{equation}
By Theorem~\ref{thm:BGG}, $K_{B,p}$ is the homology of the complex $C_B$ with terms
\begin{equation}\label{eq:finiteBGG}
C_{B,p}=(N_B\otimes V_{2p})^{\sltwo}[d_p],\qquad 0\le p\le B;
\end{equation}
there are no terms with $p>B$, since the highest weights occurring in $N_B$ are at most $2B$.

Let $\cO_B:=\C[t_1,\dots,t_B]^{S_B}$, with $\deg t_i=1$, be the coordinate ring of $\Sym^B\A^1$. Antisymmetrization identifies
\begin{equation}\label{eq:rootmodule}
N_B\cong\bigl(V_2^{\otimes B}\otimes\C[t_1,\dots,t_B]\otimes\sgn\bigr)^{S_B},\qquad
(xz^m)\cdot v=\sum_{i=1}^Bt_i^m\,x^{(i)}v,
\end{equation}
where $V_2=\sltwo$ is the adjoint representation, $x^{(i)}$ acts on the $i$-th tensor factor, and $t_i^j$ records the mode $z^j$ of the $i$-th factor. Multiplication by $\cO_B$ commutes with the action of $\lie$.

\begin{lemma}\label{lem:freeBGG}
Each $C_{B,p}$ is a finitely generated free graded $\cO_B$-module, and the differentials of $C_B$ are $\cO_B$-linear. In particular every $K_{B,p}$ is a finitely generated graded $\cO_B$-module.
\end{lemma}

\begin{proof}
By Chevalley's theorem \cite{Che55},
\begin{equation}\label{eq:Chevalley}
\C[t_1,\dots,t_B]\cong\cO_B\otimes\mathcal H_B
\end{equation}
as graded $\cO_B[S_B]$-modules, where $\mathcal H_B$ is a finite-dimensional graded $S_B$-stable complement of the ideal generated by invariants of positive degree. Since taking invariants under the finite group $S_B$ and under $\sltwo$ is exact and commutes with $\cO_B\otimes-$, we get, up to the grading shift,
\begin{equation}\label{eq:CBp-free}
C_{B,p}\cong\cO_B\otimes\bigl(V_2^{\otimes B}\otimes\mathcal H_B\otimes\sgn\otimes V_{2p}\bigr)^{S_B\times\sltwo}.
\end{equation} The differentials are induced by the $\lie$-action and the BGG maps, and therefore commute with $\cO_B$.
\end{proof}

Geometrically, $\cK_{B,p}:=\widetilde{K_{B,p}}$ is the $p$-th homology sheaf of a complex of vector bundles of length $B$ on $\Sym^B\A^1$.

\subsection{Positions}\label{subsec:positions}
The derivation $T_m$ of~\eqref{eq:T} is the transpose of the endomorphism of $N_B$ induced by multiplication by $z^m$ on $w\,\sltwo[z]$, which by~\eqref{eq:rootmodule} is multiplication by the power sum $p_m(t)=\sum_it_i^m$. Since $p_1,\dots,p_B$ generate $\cO_B$, the $\cO_B$-module structure of $K_{B,p}$ is the transpose of the action of the $T_m$ on $H_B^p$.

On the source, one checks from~\eqref{eq:XY} and~\eqref{eq:translation} that
\begin{equation}\label{eq:Tm-modes}
T_m\,\dd^k\Xc=\frac{k!}{(k-m)!}\,\dd^{k-m}\Xc,\qquad T_m\,\dd^k\Yc=3\,\frac{k!}{(k-m)!}\,\dd^{k-m}\Yc,
\end{equation}
both zero if $k<m$. On fields,
\begin{equation}\label{eq:Tm-fields}
T_m\Xc(z)=z^m\Xc(z),\qquad T_m\Yc(z)=3z^m\Yc(z).
\end{equation} Thus $T_m$ preserves $\varphi(A)$, and on $A^\vee_{r,s}\subset\C[z,y]$ the power sum $p_m(t)$ acts by multiplication by $p_m(z)+3p_m(y)$. In other words, $\cO_B$ acts on $A^\vee_{r,s}$ through the ring map
\begin{equation}\label{eq:weightedroots}
\cO_B\longrightarrow\cO_{r,s},\qquad \prod_{k=1}^B(u-t_k)\longmapsto\prod_{i=1}^r(u-z_i)\prod_{a=1}^s(u-y_a)^3,
\end{equation}
and $\psi_{B,p}$ is $\cO_B$-linear. Finally, since $\cO_B$ acts on $N_B$ by $\lie$-module endomorphisms and the comparison of Theorem~\ref{thm:BGG} is natural in the coefficient module, the identification $K_{B,p}\cong H_p(C_B)$ of \S\ref{subsec:BGG} is $\cO_B$-linear.

\subsection{The Source Sheaf}\label{subsec:source}
The ring map~\eqref{eq:weightedroots} is the comorphism of
\begin{equation}\label{eq:nu}
\nu_{r,s}:\Sym^r\A^1\times\Sym^s\A^1\longrightarrow\Sym^B\A^1,\qquad (\Pi_1(u),\Pi_3(u))\longmapsto \Pi_1(u)\,\Pi_3(u)^3,\qquad B=r+3s,
\end{equation}
where we identify effective divisors with monic polynomials. The map $\nu_{r,s}$ is finite: each $z_i$ and $y_a$ is a root of the monic polynomial on the right of~\eqref{eq:weightedroots}, whose coefficients lie in the image of $\cO_B$, so $\C[z,y]$ and its subring $\cO_{r,s}$ are finite over $\cO_B$. On the dense open set where $\Pi_1$ and $\Pi_3$ are squarefree and coprime, they are recovered from $\Pi_1\Pi_3^3$ as the products over its simple and triple roots, respectively, so $\nu_{r,s}$ is birational onto its image, the closure of the stratum of divisors with $r$ simple and $s$ triple points. Its source is smooth, so $\nu_{r,s}$ is the normalization of this closure. Since multiplication by the nonzerodivisor $F_{r,s}$ identifies $A^\vee_{r,s}$ with $\cO_{r,s}$ up to a shift of grading, we obtain
\begin{equation}\label{eq:sourcesheaf}
A^\vee_{B,p}\cong\begin{cases}\cO_{r,s}=\Gamma\bigl(\nu_{r,s*}\cO\bigr)&\text{if }(B,p)=(r+3s,r),\\ 0&\text{otherwise,}\end{cases}
\end{equation}
as ungraded $\cO_B$-modules. (We make no claim about the uniqueness of the factorization $\Pi_1\Pi_3^3$ at the boundary of the stratum.)

\subsection{The Top Degree}
\begin{lemma}\label{lem:top}
Let $B\ge1$. The top differential $C_{B,B}\to C_{B,B-1}$ vanishes, so that
\begin{equation}\label{eq:CB-split}
C_B=C_{B,B}[B]\oplus C'_B,\qquad C'_B\text{ concentrated in degrees }0,\dots,B-1,
\end{equation}
and $\psi_{B,B}$ is an isomorphism.
\end{lemma}

\begin{proof}
The $V_{2B}$-isotypic part of $V_2^{\otimes B}$ is the Cartan power, of multiplicity one and with trivial $S_B$-action. By~\eqref{eq:rootmodule}, $C_{B,B}$ is therefore isomorphic to the module of alternating polynomials $\Delta(t)\cO_B$, which is free of rank one. On the other side, $A^\vee_{B,B}=F_{B,0}\cO_{B,0}=\Delta(z)^2\C[z]^{S_B}$ is free of rank one over $\cO_B$. Since $C_{B,B}$ is the top term, $K_{B,B}$ is the kernel of the top differential, and by Proposition~\ref{prop:injective} it surjects onto $A^\vee_{B,B}$, so $K_{B,B}$ has rank one, like $C_{B,B}$, over the fraction field of the domain $\cO_B$. The image of the top differential thus has rank zero. Being a submodule of a free module it is torsion-free, hence zero. Now $\psi_{B,B}$ is a surjection between free modules of rank one, and its kernel, having rank zero and lying in a free module, vanishes.
\end{proof}

\section{Separation: Factorization at Disjoint Supports}\label{sec:separation}

We now come to the first of the three steps outlined in \S\ref{subsec:idea}, namely factorization at disjoint supports. In this section, cochains with finite-dimensional coefficients are always the full-$\Hom$ cochains of \S\ref{sec:inputs}. We return to the polynomial families of \S\ref{sec:sheaves} only through the finite free BGG complexes.

\subsection{Cup Products at Disjoint Supports}
The annihilator in $\lie$ of a finite-dimensional $\lie$-module $M$ is an ideal of finite codimension, hence of the form $\sltwo\otimes(f)$ for a monic polynomial $f$. We call the set of roots of $f$ the \emph{support} of $M$.

\begin{lemma}\label{lem:cup}
Let $M_1$ and $M_2$ be finite-dimensional $\lie$-modules with disjoint supports. Then the cup product
\begin{equation}\label{eq:cup}
C^*(\lie,\sltwo;M_1)\otimes C^*(\lie,\sltwo;M_2)\longrightarrow C^*(\lie,\sltwo;M_1\otimes M_2)
\end{equation}
is a quasi-isomorphism.
\end{lemma}

\begin{proof}
\emph{Step 1: explicit cocycles.} Use the standard triple $E,h,F$ with $[h,E]=2E$, and let $x_j\in(\lie/\sltwo)^\vee$ be the coordinate dual to $Fz^j$, $j\ge1$, with respect to the basis of modes. It has weight $2$. Let $\omega_p:=x_1\wedge\cdots\wedge x_p$. The relative differential of $x_j$ is a combination of the products $(hz^i)^\vee\wedge x_{j-i}$, $1\le i<j$, and each such term repeats a factor already present in $\omega_p$. Thus $\omega_p$ is closed. It is a highest-weight vector of weight $2p$, so it generates a copy of $V_{2p}$. Contracting this copy with the coefficient module $V_{2p}(0)$, on which $z\,\sltwo[z]$ acts by zero, gives an invariant cocycle $\alpha_{p,0}\in C^p(\lie,\sltwo;V_{2p}(0))$. Transporting it by the automorphism $z\mapsto z-a$ of $\lie$, which fixes $\sltwo$, gives a cocycle $\alpha_{p,a}$ with coefficients in $V_{2p}(a)$.

\emph{Step 2: products at distinct points are nonzero.} Let $a_1,\dots,a_n$ be distinct and $M=\bigotimes_iV_{2p_i}(a_i)$. Restrict cochains to the pair $(\hh[z],\hh)$ and decompose $M=\bigoplus_\mu M_\mu$ according to the tuples $\mu=(\mu_i)$ of weights of the factors. Each $M_\mu$ is preserved by $z\hh[z]$, which acts through the character
\begin{equation}\label{eq:character}
\chi_\mu(hf)=\sum_i\mu_if(a_i).
\end{equation}
Since $z\hh[z]$ is abelian, the restricted complex is
\begin{equation}\label{eq:restricted-complex}
\bigoplus_{\sum\mu_i=0}\Hom\bigl(\Lambda^\bullet z\hh[z],M_\mu\bigr),\qquad d\alpha=\chi_\mu\wedge\alpha;
\end{equation}
in particular the summand with $\mu=0$ is a direct summand with zero differential. Under restriction and projection to this summand, $\alpha_{p,a}$ becomes a nonzero multiple of
\begin{equation}\label{eq:ell}
\ell_{a,1}\wedge\cdots\wedge\ell_{a,p},\qquad \ell_{a,j}(hf):=\frac{f^{(j)}(a)}{j!}.
\end{equation}
Indeed, only the weight-zero vector of the copy of $V_{2p}$ generated by $\omega_p$ pairs with the zero-weight line of $V_{2p}$, and this vector is proportional to $F^p\omega_p$. The component of $F^p\omega_p$ involving only Cartan modes arises by lowering each factor $x_j$ exactly once, and the $p!$ such terms contribute with equal nonzero coefficients. Consequently the product $\prod_i\alpha_{p_i,a_i}$ projects to a nonzero multiple of $\bigwedge_{i,j}\ell_{a_i,j}$, $1\le j\le p_i$. By Hermite interpolation the functionals $\ell_{a_i,j}$, $0\le j\le p_i$, are linearly independent on $\C[z]$. Those with $j\ge1$ vanish on constants and are therefore independent on $z\C[z]$, so their wedge product is nonzero. Therefore the class of $\prod_i\alpha_{p_i,a_i}$ is nonzero.

\emph{Step 3: simple modules.} The quotient $\lie/\sltwo\otimes(f)$ is a product of jet algebras $\sltwo\otimes\C[z]/(z-a)^k$ over the roots $a$ of $f$, so a simple finite-dimensional $\lie$-module is a tensor product, over the points of its support, of simple modules of these jet algebras. On a simple module $M$ of a jet algebra the radical $\mathfrak r=\sltwo\otimes(z-a)\C[z]/(z-a)^k$ acts trivially. Indeed, since $\mathfrak r$ is nilpotent, $M$ is the direct sum of its generalized weight spaces for finitely many weights $\lambda:\mathfrak r\to\C$, which vanish on $[\mathfrak r,\mathfrak r]$ \cite[Chapter~II]{Jac62}. The connected group $SL_2$ permutes this finite set and therefore fixes each weight, so $\lambda([\sltwo,\mathfrak r])=0$. As $[\sltwo,\mathfrak r]=\mathfrak r$, all weights vanish, $\mathfrak r$ acts nilpotently, and by Engel's theorem the joint kernel of $\mathfrak r$ is a nonzero submodule, hence all of $M$. Thus simple finite-dimensional $\lie$-modules are of the form $\bigotimes_iV_{m_i}(a_i)$ with distinct $a_i$. Now let $M_1$ and $M_2$ be simple with disjoint supports. If some $m_i$ is odd, then by Theorem~\ref{thm:eval} the target and one of the two factors of the source have vanishing cohomology. Otherwise, by Theorem~\ref{thm:eval} the cohomologies of $M_1$, $M_2$ and $M_1\otimes M_2$ are one-dimensional, by Step~2 they are spanned by the products of the cocycles $\alpha_{p_i,a_i}$, and the cup product maps the generator of the source, which is the class of a product, to the generator of the target. In both cases the cup product is an isomorphism on cohomology. By the K\"unneth formula over $\C$, it is a quasi-isomorphism.

\emph{Step 4: d\'evissage.} Short exact sequences of coefficient modules split over the constant subalgebra $\sltwo$, so $C^*(\lie,\sltwo;-)$ takes them to short exact sequences of complexes. The same holds for $-\otimes M_2$ and $-\otimes_\C C^*(\lie,\sltwo;M_2)$. Since the cup product is natural, induction on the lengths of composition series of $M_1$ and $M_2$, together with the five lemma applied to the long exact cohomology sequences, extends Step~3 to arbitrary $M_1$ and $M_2$ with disjoint supports.
\end{proof}

\subsection{Coprime Splitting}
The comparison for polynomial families rests on the following version of the Chinese remainder theorem, which treats simple roots and weighted clusters of roots simultaneously and holds over nonreduced rings.

\begin{lemma}\label{lem:split}
Let $R$ be a commutative ring and let $f_1,f_2\in R[u]$ be monic of positive degrees $d_1,d_2$ with invertible resultant. Let $m_1,\dots,m_n$ be positive integers with $\sum_jm_j=d_1+d_2$. Let $\mathcal R$ be the $R$-algebra generated by $x_1,\dots,x_n$ subject to the coefficient identities of
\begin{equation}\label{eq:split-identity}
\prod_{j=1}^n(u-x_j)^{m_j}=f_1(u)f_2(u),
\end{equation}
and for $J\subseteq\{1,\dots,n\}$ let $\mathcal R_J$ be the $R$-algebra defined by
\begin{equation}\label{eq:split-RJ}
\prod_{j\in J}(u-x_j)^{m_j}=f_1(u),\qquad\prod_{j\notin J}(u-x_j)^{m_j}=f_2(u).
\end{equation}
Then the natural map
\begin{equation}\label{eq:split-iso}
\mathcal R\longrightarrow\prod_{J:\ \sum_{j\in J}m_j=d_1}\mathcal R_J
\end{equation}
is an isomorphism.
\end{lemma}

\begin{proof}
Choose $\alpha,\beta\in R[u]$ with $\alpha f_1+\beta f_2=1$. In $\mathcal R$ we have $(f_1f_2)(x_j)=0$, so the elements $e_j:=\beta(x_j)f_2(x_j)$ are idempotents:
\begin{equation}\label{eq:idempotent}
e_j(1-e_j)=(\alpha\beta f_1f_2)(x_j)=0 .
\end{equation} They split $\mathcal R$ into components indexed by the subsets $J$ on which $e_j=1$. In the component indexed by $J$, for $j\in J$ the element $f_2(x_j)$ is invertible and hence $f_1(x_j)=0$, while for $j\notin J$ the element $f_1(x_j)$ is invertible and $f_2(x_j)=0$. Put
\begin{equation}\label{eq:g12}
g_1=\prod_{j\in J}(u-x_j)^{m_j},\qquad g_2=\prod_{j\notin J}(u-x_j)^{m_j}.
\end{equation} For $j\notin J$, the identity $f_1(u)-f_1(x_j)=(u-x_j)q_j(u)$ shows that $u-x_j$ is invertible modulo $f_1$, with inverse $-q_j(u)/f_1(x_j)$. Thus $g_2$ is invertible modulo $f_1$, and $g_1g_2=f_1f_2$ implies $f_1\mid g_1$. Symmetrically $f_2\mid g_2$. On a nonzero component, division by a monic polynomial gives $\deg g_i\ge d_i$. As $\deg g_1+\deg g_2=d_1+d_2$, both are equalities, so $g_i=f_i$, and components with $\sum_{j\in J}m_j\ne d_1$ vanish. Conversely, the identities $g_i=f_i$ force the prescribed values of the idempotents. This identifies the components with the algebras $\mathcal R_J$.
\end{proof}

\subsection{Factorization}
Fix $B=B_1+B_2$ with $B_1,B_2\ge1$. Let $f_i$ be the universal monic polynomial of degree $B_i$ over $\cO_{B_i}$, and put
\begin{equation}\label{eq:chart}
\cO^\circ_{B_1,B_2}:=(\cO_{B_1}\otimes_\C\cO_{B_2})\bigl[\Res(f_1,f_2)^{-1}\bigr],\qquad U_{B_1,B_2}:=\Spec\cO^\circ_{B_1,B_2},
\end{equation}
the space of pairs of divisors with disjoint supports. The map $U_{B_1,B_2}\to\Sym^B\A^1$, $(f_1,f_2)\mapsto f_1f_2$, is \'etale: its tangent map $(h_1,h_2)\mapsto h_1f_2+f_1h_2$, with $\deg h_i<B_i$, is an isomorphism when $f_1$ and $f_2$ are coprime. In particular $\cO^\circ_{B_1,B_2}$ is flat over $\cO_B$.

Applying Lemma~\ref{lem:split} with all $m_j=1$ to $\C[t_1,\dots,t_B]\otimes_{\cO_B}\cO^\circ_{B_1,B_2}$ decomposes it into the product, over the subsets of $B_1$ labels, of the algebras of ordered roots of $f_1$ and of $f_2$. The group $S_B$ permutes these factors transitively, with stabilizer $S_{B_1}\times S_{B_2}$. Since finite-group invariants commute with base change (averaging is a projection), taking sign-twisted invariants in~\eqref{eq:rootmodule} gives an isomorphism of $\lie$-modules
\begin{equation}\label{eq:Nfactor}
N_B\otimes_{\cO_B}\cO^\circ_{B_1,B_2}\cong\bigl(N_{B_1}\otimes_\C N_{B_2}\bigr)\bigl[\Res^{-1}\bigr].
\end{equation}
At a closed point $(f_1,f_2)$ of $U_{B_1,B_2}$, the fiber of $N_{B_i}$ is annihilated by $\sltwo\otimes(f_i)$, because $f_i(t_k)=0$ in the corresponding algebra of ordered roots, so the two factors of~\eqref{eq:Nfactor} have disjoint supports in every closed fiber.

\begin{proposition}\label{prop:factor}
On $U_{B_1,B_2}$ there are isomorphisms of $\cO^\circ_{B_1,B_2}$-modules
\begin{align}
K_{B,p}\otimes_{\cO_B}\cO^\circ_{B_1,B_2}&\cong\bigoplus_{p_1+p_2=p}\bigl(K_{B_1,p_1}\otimes_\C K_{B_2,p_2}\bigr)\bigl[\Res^{-1}\bigr],\label{eq:Kfactor}\\
A^\vee_{B,p}\otimes_{\cO_B}\cO^\circ_{B_1,B_2}&\cong\bigoplus_{p_1+p_2=p}\bigl(A^\vee_{B_1,p_1}\otimes_\C A^\vee_{B_2,p_2}\bigr)\bigl[\Res^{-1}\bigr].\label{eq:Afactor}
\end{align}
\end{proposition}

We only use these as abstract isomorphisms of modules. For the structure behind them, see Remark~\ref{rem:factorization}.

\begin{proof}
For~\eqref{eq:Kfactor} we first construct a comparison of perfect complexes. The augmented complex $P_\bullet\otimes P_\bullet\to\C$, with the diagonal action of $\lie$, is exact by the K\"unneth formula and split over $\sltwo$, so relative projectivity of $P_\bullet$ provides a diagonal $P_\bullet\to P_\bullet\otimes P_\bullet$ lifting the identity of $\C$. On each inducing space $V_{2p}$ it involves only finitely many terms. For $\lie$-modules $M_1,M_2$ it induces a cup product
\begin{equation}\label{eq:cup-P}
\Hom_\lie(P_\bullet,M_1)\otimes\Hom_\lie(P_\bullet,M_2)\longrightarrow\Hom_\lie(P_\bullet,M_1\otimes M_2),
\end{equation}
defined over any coefficient ring and compatible with base change. Comparing with the standard relative Chevalley--Eilenberg resolution, which has the usual diagonal, identifies it up to homotopy with the cup product of Lemma~\ref{lem:cup}. Apply this to the $\cO$-linear duals of the finite free coefficient modules in~\eqref{eq:Nfactor}, and dualize the resulting bounded complexes of finite free $\cO^\circ_{B_1,B_2}$-modules. Since $\Hom_\lie(P_p,N^\vee)=\Hom_{\sltwo}(V_{2p},N^\vee)$ is the dual of $(N\otimes V_{2p})^{\sltwo}$, this produces a map of complexes
\begin{equation}\label{eq:Cfactor}
C_B\otimes_{\cO_B}\cO^\circ_{B_1,B_2}\longrightarrow\bigl(C_{B_1}\otimes_\C C_{B_2}\bigr)\bigl[\Res^{-1}\bigr].
\end{equation}
At a closed point, the fiber of~\eqref{eq:Cfactor} is the dual of the cup product for the finite-dimensional modules $M_i=(N_{B_i}\otimes_{\cO_{B_i}}\C_{f_i})^\vee$, where $\C_{f_i}$ is the residue field of $\cO_{B_i}$ at the point $f_i$. These have disjoint supports, and by Theorem~\ref{thm:BGG} and Lemma~\ref{lem:cup}, it is a quasi-isomorphism. The cone of~\eqref{eq:Cfactor} is a bounded complex of finite free modules whose fibers at all closed points are acyclic. Such a complex is acyclic: its homology is supported on a closed set, which, if nonempty, contains a closed point $\mathfrak m$, and after localizing at $\mathfrak m$ and cancelling isomorphisms the complex becomes a nonzero minimal complex, whose fiber at $\mathfrak m$ has zero differential and is nonzero. Thus~\eqref{eq:Cfactor} is a quasi-isomorphism. Taking homology, using flatness on the left and the K\"unneth formula over $\C$ followed by localization on the right, gives~\eqref{eq:Kfactor}.

For~\eqref{eq:Afactor}, let $(B,p)=(r+3s,r)$, since otherwise both sides vanish. By~\eqref{eq:weightedroots}, the algebra $\C[z,y]\otimes_{\cO_B}\cO^\circ_{B_1,B_2}$ is defined by the identity
\begin{equation}\label{eq:roots-identity}
f_1(u)f_2(u)=\prod_{i=1}^r(u-z_i)\prod_{a=1}^s(u-y_a)^3 .
\end{equation}
Apply Lemma~\ref{lem:split} with weight one on the $z_i$ and weight three on the $y_a$, and take $S_r\times S_s$-invariants. The orbits of components are indexed by $r_1+r_2=r$ and $s_1+s_2=s$ with $r_i+3s_i=B_i$, and the stabilizer of a component permutes each of its four lists of roots separately. By the same averaging argument,
\begin{equation}\label{eq:Ors-factor}
\cO_{r,s}\otimes_{\cO_B}\cO^\circ_{B_1,B_2}\cong\bigoplus_{\substack{r_1+r_2=r,\ s_1+s_2=s\\ r_i+3s_i=B_i}}\bigl(\cO_{r_1,s_1}\otimes_\C\cO_{r_2,s_2}\bigr)\bigl[\Res^{-1}\bigr].
\end{equation}
By~\eqref{eq:sourcesheaf} this is~\eqref{eq:Afactor}.
\end{proof}

\begin{remark}\label{rem:factorization}
The isomorphisms of Proposition~\ref{prop:factor} are in fact part of a richer structure, which we will not need. Since the $T_m$ are derivations, and power sums are additive under addition of divisors, the transposed products of $\Obs(\sltwo)$ and $A$ make the sheaves $\cK=\bigoplus_{B,p}\cK_{B,p}$ and $\cA^\vee=\bigoplus_{B,p}\widetilde{A^\vee_{B,p}}$ on $\coprod_B\Sym^B\A^1$ into factorization coalgebras on $\A^1$. That is, they are coalgebras for the convolution product along addition of divisors (graded-cocommutative for the parity of the ghost number), and their comultiplications restrict to~\eqref{eq:Kfactor} and~\eqref{eq:Afactor} over $U_{B_1,B_2}$. Moreover, $\psi$ is a morphism of factorization coalgebras.
\end{remark}

\begin{remark}\label{rem:dualA}
This point of view also gives a simple description of $\cA^\vee$. Let $\mathcal P$ be the factorization coalgebra on $\A^1$ cogenerated by an even simple point and an odd triple point, with collision orders $2$, $2$ and $3$ for collisions of two simple points, of a simple and a triple point, and of two triple points, respectively. Then Lemma~\ref{lem:dualA} is nothing but the statement that $\cA^\vee\cong\mathcal P$, and Theorem~\ref{thm:main} says that $\psi$ identifies $\cK$ with $\mathcal P$. The part of $\mathcal P$ with simple points only is the functional model of Stoyanovsky and Feigin for the relation $X(z)^2=0$ \cite{SF94}.
\end{remark}

\section{Collision: Depth along the Small Diagonal}\label{sec:collision}

Combined with induction on $B$, the factorization of the previous section shows that the kernel of $\psi_B$ is supported on the small diagonal, where all $b$-ghosts collide (see \S\ref{sec:proof}). Let us now show, using a depth argument, that it vanishes in degrees $2\le p\le B-1$. (The top degree $p=B$ was treated in Lemma~\ref{lem:top}.)

\subsection{The Small Diagonal}
For $B\ge1$, let $e_i$ denote the elementary symmetric functions of $t_1,\dots,t_B$, and let $\pp_B\subset\cO_B$ be the ideal
\begin{equation}\label{eq:collision}
\pp_B:=\Bigl(e_i-\tbinom Bi\bigl(e_1/B\bigr)^i:\ 2\le i\le B\Bigr).
\end{equation}
Its zero set is the small diagonal $Z_B$ of divisors of the form $B\cdot a$, i.e.\ of polynomials $(u-a)^B$. Its generators are coordinates on $\cO_B=\C[e_1,\dots,e_B]$ after a triangular change of variables, so they form a regular sequence of length $c=B-1$, and $\pp_B$ is prime. (For $B=1$, $\pp_1=0$.)

Recall that for a finitely generated module $M$ over a Noetherian ring $R$ and an ideal $I$, $\depth_IM$ is the supremum of the lengths of $M$-regular sequences in $I$, with $\depth_IM=\infty$ if $IM=M$ \cite[Definition~10.72.1, Tag~00LI]{Stacks}. We use the following facts.
\begin{enumerate}[label=(D\arabic*),leftmargin=3em]
\item The depth lemma for short exact sequences \cite[Section~47.11]{Stacks}, \cite[\S1.2]{BH93}.
\item $\depth_IM=\inf\{\depth M_{\mathfrak q}:\ \mathfrak q\supseteq I\}$ \cite[\S1.2]{BH93}.
\item $\depth_IM=\depth_{\sqrt I}M$, which follows from (D2).
\item If $R\to R'$ is a finite map of Noetherian rings and $M$ is a finitely generated $R'$-module, then $\depth_IM=\depth_{IR'}M$.
\end{enumerate}
To prove (D4), apply \cite[Lemma~10.72.11, Tag~0AUK]{Stacks}, which concerns a local base ring, to the finite maps $R_{\mathfrak q}\to R'_{\mathfrak q}$ for primes $\mathfrak q\supseteq I$, whose maximal ideals are the primes of $R'$ lying over $\mathfrak q$, and combine with (D2) for $R$ and for $R'$.

\subsection{A Variant of the Acyclicity Lemma}
The following is a variant of the acyclicity lemma of Peskine and Szpiro \cite[Lemme~1.8]{PS73} (see also \cite[Tag~00N0]{Stacks} and \cite{BE73}).

\begin{lemma}\label{lem:acyclic}
Let $R$ be Noetherian, let $I\subset R$ be generated by an $R$-regular sequence of length $c$, and let $C_c\to\cdots\to C_1\to C_0$ be a complex of finitely generated free $R$-modules. Suppose that for $2\le p\le c$ there are surjections $H_p(C)\twoheadrightarrow J_p$ whose kernels $M_p$ are supported in $V(I)$, and that
\begin{equation}\label{eq:depth-hyp}
\depth_IJ_p\ge p\qquad(2\le p\le c).
\end{equation}
Then $M_p=0$ for $2\le p\le c$.
\end{lemma}

\begin{proof}
Write $\overline C_p:=\operatorname{coker}(C_{p+1}\to C_p)$ and $\mathcal I_p:=\im(C_p\to C_{p-1})$. We show by descending induction on $p$, from $p=c$ to $p=2$, that $\depth_I\overline C_p\ge p-1$ and $M_p=0$. Initially $\overline C_c=C_c$ is free, so $\depth_I\overline C_c\ge c$. Suppose $\depth_I\overline C_p\ge p-1\ge1$. Since $M_p\subseteq H_p(C)\subseteq\overline C_p$ is finitely generated and supported in $V(I)$, some power $I^N$ annihilates it. Either $I$ contains a nonzerodivisor $x$ on $\overline C_p$, and then $x^N$ kills $M_p$ and acts injectively on $\overline C_p$, or $I\overline C_p=\overline C_p$, and then $\overline C_p$ vanishes at every prime containing $I$. In both cases $M_p=0$, so $H_p(C)=J_p$. The exact sequences
\begin{equation}\label{eq:acyclic-ses}
0\to H_p(C)\to\overline C_p\to\mathcal I_p\to0,\qquad 0\to\mathcal I_p\to C_{p-1}\to\overline C_{p-1}\to0
\end{equation}
and (D1) give
\begin{equation}\label{eq:depth-ineq}
\depth_I\mathcal I_p\ge\min(p-1,\,p-1)=p-1,\qquad \depth_I\overline C_{p-1}\ge\min(c,\,p-2)=p-2,
\end{equation}
which completes the induction.
\end{proof}

\begin{lemma}\label{lem:sourcedepth}
For $0<p<B$ we have $\depth_{\pp_B}A^\vee_{B,p}\ge p$.
\end{lemma}

\begin{proof}
If $A^\vee_{B,p}=0$ there is nothing to prove. Otherwise $(B,p)=(r+3s,r)$ with $s\ge1$, and by~\eqref{eq:sourcesheaf} $A^\vee_{B,p}\cong\cO_{r,s}$ as an $\cO_B$-module. Over a point of $Z_B$, the condition
\begin{equation}\label{eq:collision-condition}
\Pi_1(u)\,\Pi_3(u)^3=(u-a_0)^B
\end{equation}
forces $\Pi_1=(u-a_0)^r$ and $\Pi_3=(u-a_0)^s$, so the radical of $\pp_B\cO_{r,s}$ is the ideal of the line
\begin{equation}\label{eq:collision-line}
\bigl\{\bigl((u-a_0)^r,(u-a_0)^s\bigr):\ a_0\in\A^1\bigr\}\subset\Sym^r\A^1\times\Sym^s\A^1\cong\A^{r+s}.
\end{equation}
This line is the graph of a polynomial map over the coordinate $a_0=e_1(y)/s$, so its ideal is generated by $r+s-1$ of the coordinates of a polynomial coordinate system on $\A^{r+s}$, which form a regular sequence. By (D4) and (D3),
\begin{equation}\label{eq:source-depth}
\depth_{\pp_B}A^\vee_{B,p}=\depth_{\pp_B\cO_{r,s}}\cO_{r,s}=\depth_{\sqrt{\pp_B\cO_{r,s}}}\cO_{r,s}\ge r+s-1\ge r=p. \qedhere
\end{equation}
\end{proof}

\begin{remark}\label{rem:top}
This explains why the top term had to be split off in Lemma~\ref{lem:top}. For the pure-$X$ sector $s=0$ one has $p=r=B$, but only $\depth_{\pp_B}A^\vee_{B,B}=B-1$. Removing this sector shortens the free complex to length $B-1$, the length of the regular sequence generating $\pp_B$, and leaves the depth bounds needed in Lemma~\ref{lem:acyclic}.
\end{remark}

\section{Deformation: The Two Lowest Degrees}\label{sec:deformation}

The depth argument of the previous section gives no information in the two lowest degrees $p=0,1$. For these we deform the coefficient algebra $\C[z,w]/(w^2)$, as outlined in \S\ref{subsec:idea}.

\subsection{Two Deformations of the Square-Zero Algebra}
For $\ell\in\{0,1\}$ define odd derivations of the cochain algebra by
\begin{equation}\label{eq:deltas}
\delta_\ell b_j^a=0,\qquad \delta_\ell c_n^a=\frac12\sum_{\substack{i,k\ge0\\ i+k=n-\ell}}\epsilon^{abc}\,b_i^bb_k^c,
\end{equation}
with empty sums equal to zero, and put $D_\ell:=Q+\delta_\ell$. Each $\delta_\ell$ raises $B$ by $2$ and lowers $p$ by $1$, while $\delta_1$ raises the spin by $1$ and $\delta_0$ by $2$. One checks directly that $[\delta_1,\dd]=\delta_0$ and $[\delta_0,\dd]=0$.

\begin{lemma}\label{lem:deformed}
\begin{enumerate}[label=(\roman*),leftmargin=2em]
\item On $\cC$ we have $\delta_\ell^2=0$ and $Q\delta_\ell+\delta_\ell Q=0$.
\item $(\cC,D_1)$ computes $H^*(\sltwo[z,w]/(w^2-z),\sltwo)\cong H^*(\sltwo[u],\sltwo)$, via $z=u^2$ and $w=u$. The $u$-degree is $\wt_1:=2S-B$, and the cohomology is $\C$, in $\wt_1=0$.
\item $(\cC,D_0)$ computes $H^*(\sltwo[z,w]/(w^2-1),\sltwo)\cong H^*(\sltwo[z]\oplus\sltwo[z],\sltwo^{\rm diag})$, via $w\mapsto\pm1$. The $z$-degree is $\wt_0:=S-B$, and the cohomology is $\C\oplus\C\,\Yc$, in $\wt_0=0$.
\end{enumerate}
\end{lemma}

\begin{proof}
The algebra $\C[z,w]/(w^2-z^\ell)$ has the same basis of modes $z^n$, $wz^j$ as $\C[z,w]/(w^2)$, so $\cC$ is also the space of basic polynomial-mode cochains of $\sltwo[z,w]/(w^2-z^\ell)$ relative to $\sltwo\otimes1$. The only new brackets are
\begin{equation}\label{eq:new-brackets}
[wxz^i,wyz^j]=[x,y]z^{i+j+\ell},
\end{equation}
whose dual is the term~\eqref{eq:deltas}. Thus $D_\ell$ is the Chevalley--Eilenberg differential of $\sltwo[z,w]/(w^2-z^\ell)$ on basic cochains, in the conventions of~\eqref{eq:Q}. Since $D_\ell^2=0$ there and $Q$, $\delta_\ell$ have $B$-degrees $0$ and $2$, (i) follows by separating $B$-degrees. For $\ell=0$ this relies on basicness: on the full exterior algebra one finds
\begin{equation}\label{eq:D0-squared}
D_0^2=\tfrac12\,\epsilon_{abc}\,b_0^bb_0^c\,L_a,
\end{equation}
which vanishes on invariants.

For $\ell=1$, the isomorphism $\C[z,w]/(w^2-z)\cong\C[u]$ is
\begin{equation}\label{eq:parabola-iso}
z^n\longmapsto u^{2n},\qquad wz^j\longmapsto u^{2j+1},
\end{equation}
so the $u$-degree of $c_n^a$ and $b_j^a$ is $2n=2S-B$ and $2j+1=2S-B$, respectively. The cohomology is given by~\eqref{eq:trivial} in the variable $u$.

For $\ell=0$, $\C[z,w]/(w^2-1)\cong\C[z]\times\C[z]$ via $w\mapsto(1,-1)$, the constants $\sltwo\otimes1$ become the diagonal, and the $z$-degree of $c_n^a$ and $b_j^a$ is $n=S-B$ and $j=S-B$. By Lemma~\ref{lem:HS} applied to $\sltwo[z]=\sltwo\ltimes z\,\sltwo[z]$, by~\eqref{eq:trivial} and by the K\"unneth formula,
\begin{equation}\label{eq:pair-of-lines-coh}
H^*(\sltwo[z];\C)=H^*(\sltwo;\C),\qquad H^*\bigl(\sltwo[z]\oplus\sltwo[z];\C\bigr)=H^*(\sltwo;\C)^{\otimes2},
\end{equation}
both concentrated in $z$-degree zero. The homomorphism $(f,g)\mapsto(f(0),f(0))$ is a retraction onto $\sltwo^{\rm diag}$, so $\sltwo[z]\oplus\sltwo[z]=\sltwo^{\rm diag}\ltimes\mathfrak n$, and Lemma~\ref{lem:HS}, which respects the $z$-degree, shows that the relative cohomology is two-dimensional and concentrated in $z$-degree zero. In $\wt_0=0$ the basic cochains are $\Lambda(b_0^1,b_0^2,b_0^3)^{\sltwo}=\C\oplus\C\,\Yc$, with zero differential.
\end{proof}

\subsection{The Deformations on \texorpdfstring{$A$}{A}}
Write $X_j=\dd^jX$, $Y_j=\dd^jY$, and similarly $\Xc_j=\dd^j\Xc$, $\Yc_j=\dd^j\Yc$. With $b^a(\tau)=\sum_jb_j^a\tau^j$, $c^a(\tau)=\sum_{n\ge1}c_n^a\tau^n$ and $P^a(\tau):=\frac12\epsilon^{abc}b^b(\tau)b^c(\tau)$, the definition~\eqref{eq:deltas} reads
\begin{equation}\label{eq:delta-gen}
\delta_1c^a(\tau)=\tau P^a(\tau),\qquad \delta_0c^a(\tau)=P^a(\tau)-P^a(0).
\end{equation}
The generating functions of the $\Xc_j$ and $\Yc_j$ are
\begin{equation}\label{eq:XY-gen-tau}
\sum_j\Xc_j\frac{\tau^j}{j!}=e^{\tau\dd}\Xc=-\delta_{ab}\,b^a(\tau)\frac{d}{d\tau}c^b(\tau),\qquad e^{\tau\dd}\Yc=b^1(\tau)b^2(\tau)b^3(\tau),
\end{equation}
and with the identities
\begin{equation}\label{eq:P-identities}
\delta_{ab}\,b^aP^b=3\,b^1b^2b^3,\qquad \delta_{ab}\,b^a\frac{d}{d\tau}P^b=2\,\frac{d}{d\tau}\bigl(b^1b^2b^3\bigr)
\end{equation}
we obtain
\begin{equation}\label{eq:deltasource}
\delta_1\Xc_j=(2j+3)\,\Yc_j,\qquad \delta_0\Xc_j=2\,\Yc_{j+1},\qquad \delta_1\Yc_j=\delta_0\Yc_j=0 .
\end{equation}
Since the $\delta_\ell$ are derivations anticommuting with $Q$, they act on $H$, and by~\eqref{eq:deltasource} they preserve $\varphi(A)$. By Proposition~\ref{prop:injective} they are transported to derivations of $A$, determined by $\delta_1X_j=(2j+3)Y_j$, $\delta_0X_j=2Y_{j+1}$ and $\delta_\ell Y_j=0$.

\begin{lemma}\label{lem:homotopy}
Let $h_1,h_0$ be the odd derivations of $V$ with
\begin{equation}\label{eq:h-def}
h_1X_j=h_0X_j=0,\qquad h_1Y_j=X_j,\qquad h_0Y_j=\tfrac j2\,X_{j-1}\quad(h_0Y_0=0).
\end{equation}
They preserve $I$, hence descend to $A$, and on $A$
\begin{equation}\label{eq:contractions}
\delta_1h_1+h_1\delta_1=\wt_1=2S-B,\qquad \delta_0h_0+h_0\delta_0=\wt_0=S-B .
\end{equation}
\end{lemma}

\begin{proof}
On the four generators $X^2$, $XY$, $X\dd Y$, $Y\dd Y$ of $I$ the two derivations take the values
\begin{align}
\label{eq:h1-gens}h_1&:\ \bigl(X^2,\,XY,\,X\dd Y,\,Y\dd Y\bigr)\longmapsto\bigl(0,\,X^2,\,\tfrac12\dd(X^2),\,2X\dd Y-\dd(XY)\bigr),\\
\label{eq:h0-gens}h_0&:\ \bigl(X^2,\,XY,\,X\dd Y,\,Y\dd Y\bigr)\longmapsto\bigl(0,\,0,\,\tfrac12X^2,\,-\tfrac12XY\bigr).
\end{align}
Since $h_1$ commutes with $\dd$, it preserves the differential ideal $I$. The derivation $h_0$ satisfies $[h_0,\dd]=\frac12h_1$, so
\begin{equation}\label{eq:h0-induction}
h_0(\dd^k\rho)=\dd h_0(\dd^{k-1}\rho)+\tfrac12h_1(\dd^{k-1}\rho),
\end{equation}
which lies in $I$ for each generator $\rho$ by induction on $k$. Thus $h_0$ also preserves $I$. Both sides of each identity in~\eqref{eq:contractions} are even derivations, so it suffices to check them on generators, where they follow from~\eqref{eq:degrees} and the formulas for $\delta_\ell$. For instance,
\begin{equation}\label{eq:homotopy-check}
(\delta_1h_1+h_1\delta_1)X_j=(2j+3)X_j,\qquad(\delta_0h_0+h_0\delta_0)Y_j=jY_j .\qedhere
\end{equation}
\end{proof}

\subsection{The Lowest Two Degrees}
Regard $A$ as a subspace of $H$ via $\varphi$, and put
\begin{equation}\label{eq:Hbar}
\Hb{}_B^p:=H_B^p/\varphi(A_B^p).
\end{equation}
No algebra structure on $\Hb$ is needed.

\begin{proposition}\label{prop:low}
If $\Hb{}_b^p=0$ for all $b<B$ and all $p$, then $\Hb{}_B^0=\Hb{}_B^1=0$.
\end{proposition}

\begin{proof}
\emph{The parabola.} Fix $m>0$ and consider the subcomplex of $(\cC,D_1)$ with $\wt_1=m$. It is finite-dimensional, since every generator has positive $\wt_1$, and it is acyclic by Lemma~\ref{lem:deformed}(ii). As $\wt_1=2S-B$, the parity $\varepsilon$ of $B$ is fixed. Filter by $B$, i.e.\ by $(B-\varepsilon)/2$. The resulting spectral sequence has $E_1=H$ (at $\wt_1=m$), first differential $d_1=\delta_1$, higher differentials
\begin{equation}\label{eq:dr-parabola}
d_r:(B,p,S)\longmapsto(B+2r,\,p+1-2r,\,S+r),
\end{equation}
and converges to zero. By~\eqref{eq:deltasource},
\begin{equation}\label{eq:Hbar-ses}
0\longrightarrow\varphi(A)\longrightarrow H\longrightarrow\Hb\longrightarrow0
\end{equation}
is a short exact sequence of complexes for $d_1$, and $\varphi(A)$ is $d_1$-acyclic at $\wt_1=m>0$ by Lemma~\ref{lem:homotopy}. Thus $E_2=H(\Hb,\delta_1)$, and by hypothesis $E_2$ vanishes in all $B$-numbers below $B$.

At $(B,0)$ the incoming $d_1$ starts at $\Hb{}^1_{B-2}=0$ and the outgoing $d_1$ lands in $p=-1$, so $E_2=\Hb{}_B^0$. For $r\ge2$, outgoing differentials land in negative $p$ and incoming ones start at $B$-numbers below $B$. So $\Hb{}_B^0$ survives to $E_\infty=0$. Since this holds for every $m>0$, and $\wt_1=0$ contains only the constants, $\Hb{}_B^0=0$.

To treat $\Hb{}_B^1$, put
\begin{equation}\label{eq:VW}
\mathcal V_S:=(\Hb{}_B^1)_S,\qquad \mathcal W_S:=(\Hb{}_{B+2}^0)_S .
\end{equation}
At the positions $(B,1,S)$ and $(B+2,0,S+1)$, which have the same $\wt_1$, the $E_2$-terms are the kernel and the cokernel of $\delta_1:\mathcal V_S\to\mathcal W_{S+1}$. For $r\ge2$, all outgoing differentials land in negative $p$, and all incoming ones start at $B$-numbers at most $B-2$. Both entries therefore survive to $E_\infty=0$, and
\begin{equation}\label{eq:firstshift}
\delta_1:\mathcal V_S\xrightarrow{\ \sim\ }\mathcal W_{S+1}\qquad\text{for all }S.
\end{equation}
(When $2S-B\le0$ both sides vanish.) This uses no information about $B$-number $B+2$.

\emph{The pair of lines.} Now fix $m>0$ and use $D_0$ in $\wt_0=S-B=m$. This subcomplex is again finite-dimensional, because the only generators with $\wt_0=0$ are the three odd elements $b_0^a$, and it is acyclic by Lemma~\ref{lem:deformed}(iii). Unlike $\wt_1$, the weight $\wt_0$ does not fix the parity of $B$, but $D_0$ preserves it, since $Q$ preserves $B$ and $\delta_0$ raises it by $2$. We therefore split the subcomplex according to the parity $\varepsilon$ of $B$ and filter each part by $(B-\varepsilon)/2$. This gives a spectral sequence with $d_1=\delta_0$ and
\begin{equation}\label{eq:dr-lines}
d_r:(B,p,S)\longmapsto(B+2r,\,p+1-2r,\,S+2r),
\end{equation}
and Lemma~\ref{lem:homotopy} again identifies $E_2$ with $H(\Hb,\delta_0)$. The argument above gives
\begin{equation}\label{eq:secondshift}
\delta_0:\mathcal V_S\xrightarrow{\ \sim\ }\mathcal W_{S+2}\qquad\text{for all }S
\end{equation}
in $\wt_0>0$. In $\wt_0=0$ the whole complex is $\C\oplus\C\,\Yc\subset\varphi(A)$ by Lemma~\ref{lem:deformed}(iii), so $\Hb$ vanishes there, and there are no cochains with $\wt_0<0$. Thus~\eqref{eq:secondshift} holds for all $S$.

\emph{Conclusion.} Suppose $\Hb{}_B^1\ne0$, and let $S$ be the smallest spin with $\mathcal V_S\ne0$. Then~\eqref{eq:firstshift} and~\eqref{eq:secondshift} give
\begin{equation}\label{eq:contradiction}
0\ne\mathcal V_S\cong\mathcal W_{S+1}\cong\mathcal V_{S-1}=0,
\end{equation}
a contradiction. Therefore $\Hb{}_B^1=0$. All filtrations used are finite in each fixed $\wt_\ell$, so no convergence issues arise.
\end{proof}

\begin{remark}
Either deformation alone would allow a pair of excess classes in $(B,1)$ and $(B+2,0)$, cancelling each other in the spectral sequence. What rules them out is that the parabola and the pair of lines demand different spin shifts, $+1$ and $+2$, for the same pair, which is impossible for a grading bounded below. This replaces any Euler characteristic or residue computation in low degrees.
\end{remark}

\section{Proof of the Main Theorem}\label{sec:proof}

We can now put the three steps together.

\begin{proof}[Proof of Theorem~\ref{thm:main}]
By Proposition~\ref{prop:injective}, $\varphi$ is injective and every $\psi_{B,p}$ is surjective. We show that $\psi_{B,p}$ is an isomorphism for all $p$, by induction on $B$. For $B=0$, \eqref{eq:trivial} gives $H_0^*=\C=A_0^0$. Let $B\ge1$ and assume the claim for all smaller $B$-numbers. Equivalently, $\Hb{}_b^p=0$ for $b<B$. Write $M_p:=\ker\psi_{B,p}$.

\emph{Degrees $p=0,1$ and $p=B$.} Proposition~\ref{prop:low} gives $\Hb{}_B^0=\Hb{}_B^1=0$, i.e.\ $M_0=M_1=0$, and Lemma~\ref{lem:top} gives $M_B=0$.

\emph{Separation.} Let $B=B_1+B_2$ with $B_i\ge1$. By Proposition~\ref{prop:factor} and the induction hypothesis, $K_{B,p}\otimes_{\cO_B}\cO^\circ_{B_1,B_2}$ and $A^\vee_{B,p}\otimes_{\cO_B}\cO^\circ_{B_1,B_2}$ are isomorphic finitely generated modules over the Noetherian ring $\cO^\circ_{B_1,B_2}$. The base change of $\psi_{B,p}$ is a surjection between them. Composing it with an abstract isomorphism in the opposite direction gives a surjective endomorphism of a finitely generated module, which is injective \cite[Theorem~2.4]{Mat86}, \cite[Tag~05G8]{Stacks}. Thus the base change of $\psi_{B,p}$ is an isomorphism. By flatness, $M_p\otimes_{\cO_B}\cO^\circ_{B_1,B_2}=0$. Now let $\mathfrak m\subset\cO_B$ be a maximal ideal not containing $\pp_B$, i.e.\ a monic polynomial of degree $B$ with at least two distinct roots. Factoring it as $f_1f_2$ with $f_1,f_2$ coprime of positive degrees gives a point of some $U_{B_1,B_2}$ over $\mathfrak m$, at which the local map is flat and local, hence faithfully flat, so $(M_p)_{\mathfrak m}=0$. Since $\cO_B$ is a Jacobson ring, every nonempty locally closed subset of $\Spec\cO_B$ contains a maximal ideal. Applying this to $\Supp M_p\setminus V(\pp_B)$, we conclude
\begin{equation}\label{eq:support}
\Supp_{\cO_B}M_p\subseteq V(\pp_B)=Z_B .
\end{equation}

\emph{Collision.} By Lemma~\ref{lem:top}, $K_{B,p}=H_p(C'_B)$ for $0\le p\le B-1$, where $C'_B$ is a complex of finitely generated free $\cO_B$-modules in degrees $0,\dots,B-1$. Apply Lemma~\ref{lem:acyclic} with $R=\cO_B$, $I=\pp_B$, $c=B-1$, $J_p=A^\vee_{B,p}$ and $C=C'_B$. Its depth hypothesis is Lemma~\ref{lem:sourcedepth}. Therefore $M_p=0$ for $2\le p\le B-1$. (For $B\le2$ all degrees were already covered by the previous step.)

Thus every $\psi_{B,p}$ is an isomorphism, which completes the induction. Dualizing each finite-dimensional graded piece shows that $\varphi$ is bijective. As explained in \S\ref{subsec:phi}, $\varphi$ is compatible with products, with $\dd$ and with the $\lambda$-brackets, so it is an isomorphism of Poisson vertex algebras.
\end{proof}

\section{On Higher Rank: Large \texorpdfstring{$N$}{N} and the Nilpotence of \texorpdfstring{$X$}{X}}\label{sec:higher}

So far we have only discussed gauge group $SU(2)$, but the basic complex of \S\ref{subsec:basic} makes sense for any simple Lie algebra. For gauge group $SU(N)$, the analogue of~\eqref{eq:intro-obs} is
\begin{equation}\label{eq:obs-slN}
\Obs(\mathfrak{sl}_N):=H^*\bigl(\mathfrak{sl}_N[z,w]/(w^2),\mathfrak{sl}_N\bigr),
\end{equation}
graded by ghost number, spin and $B$-number as before. Determining this cohomology for finite $N\ge3$ seems hard, and we will not attempt it here. Instead, we determine its limit as $N\to\infty$ (\S\ref{subsec:largeN}) and prove the relation $X^N=0$ at finite $N$ (\S\ref{subsec:XN}).

For $\g=\mathfrak{gl}_N$ or $\mathfrak{sl}_N$ it is convenient to collect the coordinates into matrices. For a basis $t_\alpha$ of $\g$, with $b_j^\alpha$ and $c_n^\alpha$ the coordinates dual to $wt_\alpha z^j$ and $t_\alpha z^n$, put $b_j:=b_j^\alpha t_\alpha$ and $c_n:=c_n^\alpha t_\alpha$. These are $\g$-valued odd cochains, independent of the basis, and~\eqref{eq:CEsign} gives
\begin{equation}\label{eq:Q-matrix}
Qc_n=\sum_{i=1}^{n-1}c_ic_{n-i},\qquad Qb_j=\sum_{i=1}^{j}\bigl(c_ib_{j-i}+b_{j-i}c_i\bigr),
\end{equation}
where products are matrix products. For every $N$ we put
\begin{equation}\label{eq:XY-slN}
X:=-\operatorname{tr}(b_0c_1),\qquad Y:=\tfrac13\operatorname{tr}(b_0^3).
\end{equation}
For $N=2$, with the trace rescaled so that $\operatorname{tr}(e_ae_b)=\delta_{ab}$, these are the cochains~\eqref{eq:XY}.

\subsection{The Large \texorpdfstring{$N$}{N} Limit}\label{subsec:largeN}
Let $R=\C[z,w]/(w^2)$, $A=\C[z]$ and $\bar A=zA$. For $i\ge0$ put
\begin{equation}\label{eq:XY-i}
X^{(i)}:=-\operatorname{tr}\bigl(b_0^{2i+1}c_1\bigr),\qquad Y^{(i)}:=\frac1{2i+1}\operatorname{tr}\bigl(b_0^{2i+1}\bigr).
\end{equation}
These are cocycles, since $Qb_0=Qc_1=0$. We have $X^{(0)}=X$ and $Y^{(1)}=Y$, while $Y^{(0)}=\operatorname{tr}b_0$ vanishes for $\mathfrak{sl}_N$.

We call a cochain $\operatorname{tr}(u_1\cdots u_m)$, where each $u_s$ is one of the matrices $b_j$ or $c_n$, a \emph{single-trace cochain} of length $m$. Since all entries of $b_j$ and $c_n$ are odd,
\begin{equation}\label{eq:cyclic-relation}
\operatorname{tr}(u_1u_2\cdots u_m)=(-1)^{m-1}\operatorname{tr}(u_2\cdots u_mu_1).
\end{equation}
By~\eqref{eq:Q-matrix}, $Q$ maps a single-trace cochain of length $m$ to a sum of single-trace cochains of length $m+1$, and by~\eqref{eq:cyclic-relation} it annihilates those of length one, $\operatorname{tr}b_j$ and $\operatorname{tr}c_n$.

\begin{lemma}\label{lem:invariants}
Let $k\le N$. The basic cochains of $\mathfrak{gl}_N[z,w]/(w^2)$ of ghost number $k$ form the degree $k$ part of the free graded-commutative algebra on the single-trace cochains, subject only to the relations~\eqref{eq:cyclic-relation}. For $\mathfrak{sl}_N$ the same holds with the single-trace cochains of length at least two. These descriptions are compatible with restriction from $N+1$ to $N$.
\end{lemma}

\begin{proof}
Let $U$ be the odd vector space spanned by the symbols $b_j$ and $c_n$, and $G=GL_N$. The basic cochains of ghost number $k$ are the coinvariants of $S_k$ on $((\g^\vee)^{\otimes k})^G\otimes U^{\otimes k}$, with Koszul signs on $U^{\otimes k}$. For $\g=\mathfrak{gl}_N$ and $k\le N$, the first and second fundamental theorems of invariant theory \cite{Pro76} give a basis of $((\g^\vee)^{\otimes k})^G$, permuted by $S_k$ through conjugation,
\begin{equation}\label{eq:tau-sigma}
\tau_\sigma(x_1\otimes\cdots\otimes x_k)=\prod_{(i_1\cdots i_r)}\operatorname{tr}(x_{i_1}\cdots x_{i_r}),\qquad\sigma\in S_k,
\end{equation}
the product being over the cycles of $\sigma$. Passing to coinvariants, $\tau_\sigma\otimes u_1\otimes\cdots\otimes u_k$ becomes the product of the single-trace cochains read off from the cycles of $\sigma$, which proves the claim for $\mathfrak{gl}_N$. For $\mathfrak{sl}_N$ the restricted $\tau_\sigma$ span, and those for which $\sigma$ has a fixed point vanish. The others are linearly independent, by induction on $k$ and a count of dimensions using $(\mathfrak{gl}_N^\vee)^{\otimes k}=\bigoplus_{S\subseteq\{1,\dots,k\}}(\mathfrak{sl}_N^\vee)^{\otimes S}$. Compatibility with restriction is clear.
\end{proof}

\begin{theorem}\label{thm:stable}
In each trigraded degree, restriction along $\mathfrak{sl}_N\subset\mathfrak{sl}_{N+1}$ induces an isomorphism
\begin{equation}\label{eq:stability}
H^*\bigl(\mathfrak{sl}_{N+1}[z,w]/(w^2),\mathfrak{sl}_{N+1}\bigr)\xrightarrow{\ \sim\ }H^*\bigl(\mathfrak{sl}_N[z,w]/(w^2),\mathfrak{sl}_N\bigr)\qquad\text{in ghost numbers }\gh<N.
\end{equation}
The stable limit $H_\infty$ is the free graded-commutative algebra on the classes
\begin{equation}\label{eq:stable-generators}
\dd^aX^{(i)}\quad(i\ge0,\ a\ge0),\qquad \dd^aY^{(i)}\quad(i\ge1,\ a\ge0),
\end{equation}
of degrees $(B,p,S)=(2i+1,1,a+2i+2)$ and $(2i+1,0,a+2i+1)$, respectively. The former are even and the latter odd.
\end{theorem}

\begin{proof}
By Lemma~\ref{lem:invariants}, the complexes of basic cochains in ghost numbers at most $N$ do not depend on $N$, which gives~\eqref{eq:stability}. Let $T^\vee$ be the complex of formal single-trace cochains, that is, cyclic words in the $b_j$ and $c_n$ subject to~\eqref{eq:cyclic-relation}, with the differential~\eqref{eq:Q-matrix}. By the remarks before the lemma, $T^\vee=T^\vee_1\oplus T^\vee_{\ge2}$ as complexes, according to length, and $T^\vee_1$ has zero differential. In the stable range the basic complex of $\mathfrak{sl}_N$ is the free graded-commutative algebra on $T^\vee_{\ge2}$, so $H_\infty$ is the free graded-commutative algebra on $H(T^\vee_{\ge2})$. It therefore suffices to show that $H(T^\vee)$ has the basis
\begin{equation}\label{eq:single-trace-basis}
\operatorname{tr}c_n\ \ (n\ge1),\qquad \dd^a\operatorname{tr}(b_0^k),\quad\dd^a\operatorname{tr}(b_0^kc_1)\ \ (k\ge1\text{ odd},\ a\ge0),
\end{equation}
since the classes of length one here are $\operatorname{tr}c_n$ and $\dd^a\operatorname{tr}b_0=a!\operatorname{tr}b_a$, and the others are, up to normalization, the classes~\eqref{eq:stable-generators}.

The differential preserves the number $k$ of letters $b_j$, so $T^\vee=\bigoplus_{k\ge0}T^\vee_k$. In $B$-number $0$ the basic complex of $\mathfrak{sl}_N$ is the relative complex of $(\mathfrak{sl}_N[z],\mathfrak{sl}_N)$, whose cohomology is $\C$ \cite{GL76}. Thus $H(T^\vee_0\cap T^\vee_{\ge2})=0$, and $H(T^\vee_0)$ has the basis $\operatorname{tr}c_n$.

Let $k\ge1$. The graded dual $T_k$ of $T^\vee_k$ is spanned by cyclic words in odd letters $z^n$ ($n\ge1$) and $wz^j$ ($j\ge0$), with exactly $k$ letters of the second kind, and its differential replaces pairs of cyclically adjacent letters by their product in $R$. Cutting a cyclic word in front of each letter $wz^j$ identifies $T_k$ with the coinvariants of $\mathbb Z/k$ on the complex $L_k$ of linear words $\beta_1\gamma_1\cdots\beta_k\gamma_k$, where the $\beta_s$ are letters $wz^j$, the $\gamma_s$ are words in $\bar A$, and $\mathbb Z/k$ rotates the blocks $\beta_s\gamma_s$. With $M=wA$ and $\operatorname{Bar}(A)$ the normalized bar resolution
\begin{equation}\label{eq:bar}
\operatorname{Bar}(A)=\bigoplus_{m\ge0}A\otimes\bar A^{\otimes m}\otimes A\longrightarrow A
\end{equation}
of $A$ as a bimodule, $L_k$ is the cyclic tensor product
\begin{equation}\label{eq:Lk-cyclic}
L_k=\bigl(M\otimes_A\operatorname{Bar}(A)\otimes_AM\otimes_A\operatorname{Bar}(A)\otimes_A\cdots\otimes_AM\otimes_A\operatorname{Bar}(A)\bigr)\otimes_{A\otimes A^{\rm op}}A
\end{equation}
with $k$ factors $M$. Since $\operatorname{Bar}(A)$ consists of free bimodules, replacing all but one copy of $\operatorname{Bar}(A)$ by $A$ is a quasi-isomorphism, onto the Hochschild complex of $A$ with coefficients in $M^{\otimes_Ak}\cong A$. By the Hochschild--Kostant--Rosenberg theorem \cite{HKR62},
\begin{equation}\label{eq:HKR}
H_*(L_k)\cong HH_*(A)=A\oplus A\,dz ,
\end{equation}
with $z^a$ in degree $(B,p,S)=(k,0,k+a)$ and $z^a\,dz$ in degree $(k,1,k+a+1)$.

Since $\mathbb Z/k$ is finite, $H(T_k)=H(L_k)_{\mathbb Z/k}$, which is at most one-dimensional in each degree. The derivation $\dd_z$ of $R$ acts on all of the above, compatibly with $\mathbb Z/k$, and on~\eqref{eq:HKR} it maps $z^{a+1}$ and $z^{a+1}dz$ to $(a+1)z^a$ and $(a+1)z^a\,dz$. Dually, as $\dd$ is the transpose of $\dd_z$, $H(T^\vee_k)$ has the basis $\dd^a\phi$, $a\ge0$, where $\phi$ runs over its nonzero classes in the lowest degrees $(k,0,k)$ and $(k,1,k+1)$. Up to rotation, the only cochains there are the cocycles $\operatorname{tr}(b_0^k)$ and $\operatorname{tr}(b_0^kc_1)$. If $k$ is even, $\operatorname{tr}(b_0^k)=0$ by~\eqref{eq:cyclic-relation}, and
\begin{equation}\label{eq:trace-exact-even}
Q\operatorname{tr}\bigl(b_0^{k-1}b_1\bigr)=-\operatorname{tr}\bigl(b_0^{k-1}c_1b_0\bigr)-\operatorname{tr}\bigl(b_0^kc_1\bigr)=-2\operatorname{tr}\bigl(b_0^kc_1\bigr),
\end{equation}
so $T^\vee_k$ is acyclic. If $k=2i+1$ is odd, $\operatorname{tr}(b_0^k)\ne0$ is not exact, as it has $p=0$, and $\operatorname{tr}(b_0^kc_1)$ is not exact, as the only cochain of degree $(k,0,k+1)$ is closed:
\begin{equation}\label{eq:trace-closed}
Q\operatorname{tr}\bigl(b_0^{2i}b_1\bigr)=\operatorname{tr}\bigl(b_0^{2i}c_1b_0\bigr)+\operatorname{tr}\bigl(b_0^{2i+1}c_1\bigr)=0 .
\end{equation}
This proves~\eqref{eq:single-trace-basis}.
\end{proof}

In fact, the $\lambda$-brackets of these generators close among themselves, with coefficients that do not depend on $N$.

\begin{proposition}\label{prop:stable-brackets}
For every $N\ge2$, the classes~\eqref{eq:XY-i} in $\Obs(\mathfrak{sl}_N)$ have the $\lambda$-brackets
\begin{align}
\{X^{(i)}{}_\lambda X^{(j)}\}&=\bigl((2i+1)\dd+(2i+2j+2)\lambda\bigr)X^{(i+j)},\label{eq:XX-stable}\\
\{X^{(i)}{}_\lambda Y^{(j)}\}&=\bigl((2i+1)\dd+(2i+2j+1)\lambda\bigr)Y^{(i+j)},\label{eq:XY-stable}\\
\{Y^{(i)}{}_\lambda Y^{(j)}\}&=0.\label{eq:YY-stable}
\end{align}
Consequently $H_\infty\cong S(R_\infty)$ as Poisson vertex algebras, where
\begin{equation}\label{eq:R-infinity}
R_\infty=\bigoplus_{i\ge0}\C[\dd]X^{(i)}\oplus\bigoplus_{i\ge1}\C[\dd]Y^{(i)}
\end{equation}
is the Lie conformal superalgebra with these brackets. It contains the Lie conformal superalgebra $R$ of \S\ref{subsec:A}, spanned by $X=X^{(0)}$ and $Y=Y^{(1)}$.
\end{proposition}

\begin{proof}
As in \S\ref{subsec:phi}, write $b=b_0$ and $\dd c=c_1$. The $\lambda$-bracket of $\Obs(\mathfrak{sl}_N)$ is induced by that of the $bc$-system \cite{Kha26}, which in matrix notation reads
\begin{equation}\label{eq:bc-matrix}
\{b_{kl}{}_\lambda c_{mn}\}=\delta_{kn}\delta_{lm}-N^{-1}\delta_{kl}\delta_{mn},
\end{equation}
with $\{b_\lambda b\}=\{c_\lambda c\}=0$, so $\{b_{kl}{}_\lambda\dd c_{mn}\}$ is $\lambda$ times~\eqref{eq:bc-matrix}. The cochains~\eqref{eq:XY-i} are polynomials in the entries of $b$ and $\dd c$, so by the Leibniz rules the $\lambda$-bracket of two of them is $B_0+\lambda B_1$, where $B_1$ is given by the super-biderivation $P$ with $P(b_{kl},\dd c_{mn})=-P(\dd c_{mn},b_{kl})$ equal to~\eqref{eq:bc-matrix} and $P(b,b)=P(\dd c,\dd c)=0$. In particular~\eqref{eq:YY-stable} holds. The second term of~\eqref{eq:bc-matrix} contributes to $B_0$ and $B_1$ only products of two traces, one of which is the cocycle $\operatorname{tr}(b^{2l+1})$ or its derivative, and the other $\operatorname{tr}(b^{2l})=0$, $\operatorname{tr}(\dd c)=0$, or $\operatorname{tr}(b^{2l}\dd c)$ with $l\ge1$ or its derivative, which is exact by~\eqref{eq:trace-exact-even}. These contributions are exact, and we drop them. What remains of $B_0$ and $B_1$ are single-trace cocycles given by formulas that do not depend on $N$.

For $B_1$, the Leibniz rule now gives, as matrices,
\begin{equation}\label{eq:P-matrices}
P\bigl(X^{(i)},b\bigr)=b^{2i+1},\qquad P\bigl(X^{(i)},\dd c\bigr)=\sum_{p=1}^{2i+1}(-1)^{p+1}\,b^{2i+1-p}\,\dd c\,b^{p-1},
\end{equation}
and $P(X^{(i)},-)$ is an even derivation. Using
\begin{equation}\label{eq:rotation}
\operatorname{tr}\bigl(b^{m-p}\,\dd c\,b^{p-1}\bigr)=(-1)^{p+1}\operatorname{tr}\bigl(b^{m-1}\dd c\bigr)\qquad(m\text{ even}),
\end{equation}
which follows from~\eqref{eq:cyclic-relation}, we obtain
\begin{equation}\label{eq:P-results}
P\bigl(X^{(i)},X^{(j)}\bigr)=(2i+2j+2)X^{(i+j)},\qquad P\bigl(X^{(i)},Y^{(j)}\bigr)=(2i+2j+1)Y^{(i+j)}.
\end{equation}

Next, $B_0$ lies in degree $(2i+2j+1,1,2i+2j+3)$, respectively $(2i+2j+1,0,2i+2j+2)$. Being closed for every $N$, it is a cocycle of $T^\vee$ by Lemma~\ref{lem:invariants}, so by~\eqref{eq:single-trace-basis} it is cohomologous in $T^\vee$, and hence for every $N$, to $\alpha\,\dd X^{(i+j)}$, respectively $\alpha'\,\dd Y^{(i+j)}$, for constants $\alpha$, $\alpha'$. To find them, take $N$ large. Since $b$ and $\dd c$ are traceless, $\{X_\mu b\}=(\dd+\mu)b$ and $\{X_\mu\dd c\}=(\dd+\mu)\dd c$ as in \S\ref{subsec:phi}, so by the Leibniz rule $\{X_\mu f\}=(\dd+d\mu)f$ for every polynomial $f$ of degree $d$ in the entries of $b$ and $\dd c$. Thus $X^{(i)}$ and $Y^{(i)}$ are primary of weights $2i+2$ and $2i+1$:
\begin{equation}\label{eq:primary}
\{X_\mu X^{(i)}\}=\bigl(\dd+(2i+2)\mu\bigr)X^{(i)},\qquad \{X_\mu Y^{(i)}\}=\bigl(\dd+(2i+1)\mu\bigr)Y^{(i)} .
\end{equation}
If $f$, $g$, $h$ are primary of weights $h_f$, $h_g$, $h_h$, and $\{f_\lambda g\}=(\alpha\dd+\beta\lambda)h$ with $h\ne0$, then comparing the coefficients of $\mu^2$ in the Jacobi identity
\begin{equation}\label{eq:jacobi-primary}
\{X_\mu\{f_\lambda g\}\}=\{\{X_\mu f\}_{\lambda+\mu}g\}+\{f_\lambda\{X_\mu g\}\}
\end{equation}
gives $h_h\,\alpha=(h_f-1)\,\beta$. By~\eqref{eq:P-results} and~\eqref{eq:primary}, and since $X^{(i+j)}\ne0$ and $Y^{(i+j)}\ne0$ for $N$ large by Theorem~\ref{thm:stable}, this gives $\alpha=\alpha'=2i+1$, which proves~\eqref{eq:XX-stable} and~\eqref{eq:XY-stable}. As these brackets do not depend on $N$, and the classes~\eqref{eq:XY-i} are linearly independent for $N$ large, $R_\infty$ is a Lie conformal superalgebra, and by Theorem~\ref{thm:stable} $H_\infty$ is its symmetric algebra.
\end{proof}

\begin{corollary}\label{cor:stable-index}
The Hilbert--Poincar\'e series of $H_\infty$ is
\begin{equation}\label{eq:stable-poincare}
P_{H_\infty}(t,q,y)=\prod_{i\ge0}\frac1{(t^{2i+2}y^{2i+1}q^{2i+2};q)_\infty}\;\prod_{i\ge1}\bigl(-t^{2i+1}y^{2i+1}q^{2i+1};q\bigr)_\infty .
\end{equation}
In particular, the large $N$ Schur index is
\begin{equation}\label{eq:stable-schur}
P_{H_\infty}(-1,q,1)=\frac1{(q^2;q)_\infty}\prod_{i\ge1}\bigl(1-q^{2i+1}\bigr)=\frac1{(q^2;q^2)_\infty}.
\end{equation}
\end{corollary}

This agrees with the Schur index of the pure $SU(N)$ theory, conjectured by Okazaki and Smith \cite{OS24} and a consequence of the Macdonald identities \cite{Mac72},
\begin{equation}\label{eq:Lu}
\mathcal I_{\mathrm S}^{SU(N)}(q)=\frac{(q^{2N};q^{2N})_\infty^N}{(q^2;q^2)_\infty}=\frac1{(q^2;q^2)_\infty}-N\,q^{2N}+O\bigl(q^{2N+2}\bigr),
\end{equation}
which tends to~\eqref{eq:stable-schur} as $N\to\infty$. For $N=2$ it is~\eqref{eq:intro-Schur}. Thus the Schur index first deviates from the stable one in spin $2N$, which is the spin of $X^N$.

\subsection{The Relation \texorpdfstring{$X^N=0$}{X\^{}N=0}}\label{subsec:XN}
Let us now turn to finite $N$. The following result generalizes the relation $X^2=0$ of Theorem~\ref{thm:main}, which is the case $N=2$.

\begin{theorem}\label{prop:XN}
Let $N\ge2$. In $H^*(\mathfrak{sl}_N[z,w]/(w^2),\mathfrak{sl}_N)$ we have $X^{N-1}\ne0$ and $X^N=0$.
\end{theorem}

\begin{proof}
Let $h_1,\dots,h_{N-1}$ be a basis of a Cartan subalgebra of $\mathfrak{sl}_N$. The elements $wh_k$ and $zh_k$ of $\g=\mathfrak{sl}_N[z,w]/(w^2)$ commute, so
\begin{equation}\label{eq:cartan-cycle}
\Omega:=wh_1\wedge\cdots\wedge wh_{N-1}\wedge zh_1\wedge\cdots\wedge zh_{N-1}
\end{equation}
is a Chevalley--Eilenberg cycle and pairs to zero with every coboundary. Since $\langle X^{N-1},\Omega\rangle=\pm(N-1)!\,\det\bigl(\operatorname{tr}(h_kh_l)\bigr)\ne0$, we get $X^{N-1}\ne0$.

For the vanishing we pass to $\mathfrak{gl}_N$. Since $\mathfrak{gl}_N(R)=\mathfrak{sl}_N(R)\times R$,
\begin{equation}\label{eq:gl-sl}
H^*\bigl(\mathfrak{gl}_N(R),\mathfrak{gl}_N\bigr)=H^*\bigl(\mathfrak{sl}_N(R),\mathfrak{sl}_N\bigr)\otimes\Lambda\bigl(\bar R^\vee\bigr),\qquad \bar R=(z,w),
\end{equation}
where the second factor is generated by the central coordinates $\operatorname{tr}b_j$ and $\operatorname{tr}c_n$. The matrix $\Pi:=b_0c_1$ has commuting entries, so $e_{N+1}(\Pi)=0$, and by Newton's identities $\operatorname{tr}(\Pi)^{N+1}$ is a polynomial in the $\operatorname{tr}(\Pi^k)$ each of whose monomials has a factor with $k\ge2$. These traces are closed, as $Qb_0=Qc_1=0$, and for $k\ge2$ they are exact: with $M:=\Pi^{k-2}$,~\eqref{eq:Q-matrix} gives
\begin{equation}\label{eq:trace-exact}
\operatorname{tr}(\Pi^k)=-Q\operatorname{tr}\bigl(b_0b_1c_1M\bigr)-Q\operatorname{tr}\bigl(b_0^2c_2M\bigr).
\end{equation}
So $\operatorname{tr}(\Pi)^{N+1}$ is exact. Write $b_0=b_0'+\frac1N\beta\,\mathbf 1$ and $c_1=c_1'+\frac1N\gamma\,\mathbf 1$ with $b_0'$, $c_1'$ traceless, $\beta=\operatorname{tr}b_0$ and $\gamma=\operatorname{tr}c_1$. Then $\operatorname{tr}(\Pi)=-X+\frac1N\beta\gamma$, with $X$ the cochain~\eqref{eq:XY-slN} of $\mathfrak{sl}_N$, and as $(\beta\gamma)^2=0$,
\begin{equation}\label{eq:XN-gl}
\operatorname{tr}(\Pi)^{N+1}=(-X)^{N+1}+\frac{N+1}{N}\,(-X)^N\,\beta\gamma .
\end{equation}
Since $\beta\gamma$ is a nonzero class in the second factor of~\eqref{eq:gl-sl}, it follows that $X^N=0$.
\end{proof}

An analogous relation holds in the classical chiral ring of pure $\mathcal N=1$ super Yang--Mills theory, where the glueball superfield $S\sim\operatorname{tr}W_\alpha W^\alpha$ satisfies $S^{h^\vee}=0$, with $h^\vee$ the dual Coxeter number, so that $S^N=0$ for $SU(N)$ \cite{CDSW02} (see also \cite{Wit03,EK03}).

For $N\ge3$ we are not aware of a standing conjectural presentation of $\Obs(\mathfrak{sl}_N)$. It seems plausible that, as a differential algebra, it is generated by the classes
\begin{equation}\label{eq:finite-generators}
X^{(i)}=-\operatorname{tr}\bigl(b_0^{2i+1}c_1\bigr)\quad(0\le i\le N-2),\qquad Y^{(i)}=\frac1{2i+1}\operatorname{tr}\bigl(b_0^{2i+1}\bigr)\quad(1\le i\le N-1),
\end{equation}
but the relations appear to be intricate. It would be interesting to compute $H^*(\mathfrak{sl}_N[z,w]/(w^2),\mathfrak{sl}_N)$ for $N\ge3$, generalizing the results of this paper to higher rank.

\end{document}